\documentclass{LMCS}

\usepackage{amssymb}
\usepackage{amsmath}
\usepackage{latexsym}
\usepackage{pst-tree}
\usepackage{eufrak}
\usepackage{epic}
\usepackage{eepic}
\usepackage{hyperref}

\def\doi{2 (3:2) 2006}
\lmcsheading%
{\doi}
{1--31}
{}
{}
{Aug.~31, 2005}
{Jul.~26, 2006}
{}

\begin{document}

\title[Logics for unranked trees]{Logics for Unranked Trees: An
  Overview\rsuper*}

\author[L.~Libkin]{Leonid Libkin}       %required
\address{School of Informatics, University of Edinburgh, and Department of Computer Science, University of Toronto}     %required
\email{libkin@inf.ed.ac.uk and libkin@cs.toronto.edu}  %optional
\titlecomment{{\lsuper *}An earlier
version of this paper appeared in the Proceedings of the 32nd
International Colloquium on Automata, Languages, and Programming (ICALP
2005).}


%%%%%%%%%%%%%%%%%%%%%%%%%%%%%%%%%%%%%%%%%%%%%%%%%%%%%%%%%%%%%%%%%%%%%%%%%%%

%% the abstract has to preceed the command \maketitle:
\begin{abstract}
  \noindent Labeled unranked trees are used as a model of XML documents, and
logical languages for them have been studied actively over the past
several years. 
Such logics have different purposes: some are better suited for
extracting data, some for expressing navigational properties, and some
make it easy to relate complex properties of trees to the
existence of tree automata for those properties. Furthermore, logics
differ significantly in their model-checking properties, their
automata models, and their behavior on ordered and unordered trees. 
In this paper we present a survey of logics for unranked
trees. 
\end{abstract}

\maketitle

\newcommand{\Blue}[1]{#1}

\newcommand{\fth}{\hfill \mbox{\fbox{}}}

\newcommand{\LB}{\mathopen{\{\!|}}
\newcommand{\RB}{\mathclose{|\!\}}}
\newcommand{\BAG}[1]{\LB #1 \RB}
\newcommand{\BIN}{\mbox{ $\in\!\!\!\in$ }}

\newcommand{\bag}{\BAG}
\newcommand{\BAGK}[2]{{\BAG #1}_#2}
\newcommand{\bagk}{\BAGK}

\newcommand{\SET}[1]{\{ #1 \}}

\newcommand{\SEPR}{\hspace*{0.4cm}}

\renewcommand{\AA}{{\mathcal A}}
\newcommand{\GG}{{\mathcal G}}
\newcommand{\CC}{{\mathcal C}}
\newcommand{\LL}{{\mathcal L}}
\newcommand{\BB}{{\mathcal B}}
\newcommand{\KK}{{\mathcal K}}
\newcommand{\II}{{\mathcal I}}
\newcommand{\OO}{{\mathcal O}}
\newcommand{\TT}{{\mathcal T}}
\newcommand{\FF}{{\mathcal F}}
\newcommand{\UU}{{\mathcal U}}
\newcommand{\MM}{{\mathcal M}}
\newcommand{\SSS}{{\mathcal S}}
\newcommand{\NN}{{\mathcal N}}
\newcommand{\PP}{{\mathcal P}}
\newcommand{\DD}{{\mathcal D}}
\newcommand{\HH}{{\mathcal H}}
\newcommand{\QQ}{{\mathcal Q}}

\newcommand{\MMM}{\EuFrak{M}}
\newcommand{\TTT}{\EuFrak{T}}

\newcommand{\FO}{{\rm FO}}
\newcommand{\MSO}{{\rm MSO}}
\newcommand{\CMSO}{\MSO_{\text{\rm mod}}}
\newcommand{\PMSO}{{\rm PMSO}}

\newcommand{\lr}{{\sf lr}}
\newcommand{\hlr}{{\sf h}\lr}
\newcommand{\fsv}{\hlr}
\newcommand{\qr}{{\sf qr}}
\newcommand{\Fraisse}{{Fra\"\i ss\'e\ }}
\newcommand{\EF}{{Ehrenfeucht-Fra\"\i ss\'e\ }}

\newcommand{\LRA}{\Leftrightarrow}
\newcommand{\lra}{\leftrightarrow}

\newcommand{\LLinfty}{\LL_{\infty\omega}}
\newcommand{\LLfinvar}{\LLinfty^\omega}
\newcommand{\LLstar}{\LLinfty^\ast}
\newcommand{\LLstarcount}{\LLstar({\bf C})}
\newcommand{\ucl}{\LLstarcount}
\newcommand{\sucl}{\LLinfty^\circ({\bf C})}
\newcommand{\LLfinvarcount}{\LLfinvar({\bf C})}
\newcommand{\LLinftycount}{\LLinfty({\bf C})}
\newcommand{\ct}[2]{\# #1.#2}
\newcommand{\fv}{{\mbox{\sc Fv}}}
\newcommand{\rank}{{\sf rk}}
\newcommand{\rk}{\rank}
\newcommand{\deq}{{\leftrightarrows}}
\newcommand{\wdeq}{\deq^w}
\newcommand{\floor}[1]{\lfloor #1 \rfloor}
\newcommand{\existsbang}{\exists!}

\newcommand{\qq}{{\mathbb{Q}}}
\newcommand{\nn}{{\mathbb{N}}}
\newcommand{\zz}{{\mathbb{Z}}}
\newcommand{\rr}{{\mathbb{R}}}
\def\vec{\mathaccent"017E }
\newcommand{\crc}[1]{#1^{\circ}}
\newcommand{\e}{\varepsilon}

%%%%%%%%% Unranked tree stuff
\newcommand{\da}{\downarrow\!}
\newcommand{\precr}{\prec_{\rightarrow}}
\newcommand{\precd}{\prec_{\da}}
\newcommand{\preceqr}{\preceq_{\rightarrow}}
\newcommand{\preceqd}{\preceq_{\da}}

\newcommand{\FOeta}{\FO_\eta}
\newcommand{\FOdom}{\FO_{{\rm dom}}}
\newcommand{\actFO}{\FO^{{\rm act}}}
\newcommand{\FOetaact}{\FO^{{\rm act}}_\eta}

\newcommand{\SwS}{\text{S}\omega\text{S}}
\newcommand{\StwoS}{\text{S2S}}
\newcommand{\RT}{{\mathcal R}}

\newcommand{\oleft}{\mbox{$\sqsubset\!\!\!{}_{{}_\rightarrow}$}}
\newcommand{\odown}{\sqsubset\!\!_{{}_\downarrow}}

\newcommand{\foreg}{{\mathcal FOREG}}
\newcommand{\FOREG}{\text{${\mathcal FOREG}$}}
\newcommand{\FOreg}{\mbox{$\FOeta^{\text{reg}}$}}
\newcommand{\FOregdom}{\text{FO$_{\eta,\text{dom}}^{\text{reg}}$}}

\newcommand{\efeq}{\equiv}
\newcommand{\efeqv}{\equiv^V}
\newcommand{\efeqw}{\equiv^W}
\newcommand{\efeqffo}{\efeq^{{\scriptscriptstyle \FFO}}}
\renewcommand{\efeqffo}{\efeq}
\newcommand{\efeqeta}{\efeq^{\eta}}
\newcommand{\efeqetareg}[1]{\efeq^{\eta,#1}}
\newcommand{\efeqpath}{\efeq^{{\scriptscriptstyle \MSOpath}}}
\newcommand{\efeqmsopc}{\efeq^{\scriptscriptstyle \MSOpc}}
\newcommand{\efeqmso}{\efeq^{\scriptscriptstyle \mso}}
\newcommand{\efeqdom}{\efeq^{{\rm dom}}}
\newcommand{\efeqforeg}[1]{\efeq^{\foreg,#1}}

\newcommand{\pre}{<_{\text{pre}}}
\newcommand{\preeq}{\leq_{\text{pre}}}

\newcommand{\eb}{\exists^{\eta}}
\newcommand{\fb}{\forall^{\eta}}
\newcommand{\ep}{\exists^{\text{path}}}

\newcommand{\calcmu}{L_\mu} 
\newcommand{\muc}{\calcmu}
\newcommand{\calcmuq}{C^{{\rm mod}}_\mu}
\newcommand{\mucmod}{\calcmuq}
\newcommand{\calcmuth}{C_\mu}
\newcommand{\cmuc}{\calcmuth}
\newcommand{\calcmuthsim}{\cmuc}
\newcommand{\calcmusib}{\muc^{{\rm full}}}
\newcommand{\fmuc}{\calcmusib}

\newcommand{\bx}{\Box}
\newcommand{\dm}{\Diamond}

\newcommand{\ctl}{{\rm CTL}}
\newcommand{\ctls}{\ctl^\star}
\newcommand{\ctlspast}{\ctls_{\text{\rm past}}}

\newcommand{\F}{{\mathbf F}}
\newcommand{\X}{{\mathbf X}}
\newcommand{\Y}{{\mathbf X}^-}
\newcommand{\E}{{\mathbf E}}

\newcommand{\HP}{{\mathbf H}}
\newcommand{\EX}{\E\X}
\newcommand{\U}{{\mathbf U}}
\renewcommand{\S}{{\mathbf S}} 

\newcommand{\cctl}{\ctls_{\rm count}}

\newcommand{\tree}{\text{\sc Tree}}
\newcommand{\treeq}{\tree_{\qq}}
\newcommand{\Def}{\text{\sf Def}}
\newcommand{\el}{{\sf el}}
\newcommand{\lex}{<_{\text{\rm lex}}}
\newcommand{\dom}{{\rm dom}}
\newcommand{\domeq}{\approx_\dom}

\renewcommand{\phi}{\varphi}

 \newtheorem{example}[thm]{Example}
 \newtheorem{lemma}[thm]{Lemma}
 \newtheorem{proposition}[thm]{Proposition}
 \newtheorem{corollary}[thm]{Corollary}

\psset{levelsep=0.7cm,treemode=D, nodesepB=2pt,nodesepA=2pt}

\parindent=0cm
\parskip=0.27cm

\section{Introduction}

Trees arise everywhere in computer science, and there are numerous
formalisms in the literature for describing and manipulating trees.
Some of these formalisms are declarative and based on logical
specifications: for example, first-order logic,  monadic
second-order logic, and various temporal and fixed-point logics over
trees. Others are procedural formalisms such as
various flavors of tree automata, or tree transducers, or tree
grammars.
All these formalisms have found numerous applications in verification,
program analysis, logic programming, constraint programming,
linguistics, and databases.

\begin{figure}
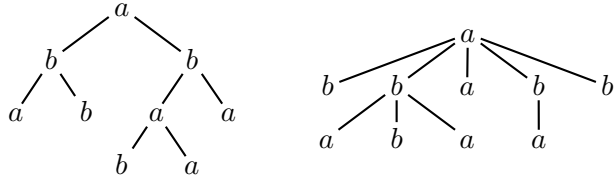
\rm
\begin{center}
\begin{minipage}{4cm}
\pstree{\TR{$a$}}{\pstree{\TR{$b$}}{\TR{$a$}\TR{$b$}}
{\pstree{\TR{$b$}}{\pstree{\TR{$a$}}{\TR{$b$}\TR{$a$}}\TR{$a$}}}}
\end{minipage}
\begin{minipage}{4cm}
\pstree{\TR{$a$}}{\TR{$b$}\pstree{\TR{$b$}}{\TR{$a$}\TR{$b$}\TR{$a$}}\TR{$a$}
\pstree{\TR{$b$}}{\TR{$a$}}\TR{$b$}}
\end{minipage}
\caption{\label{fig:example} {\rm A ranked (binary) and an unranked tree}}
\end{center}
\end{figure}

Until recently, most logical formalisms for trees dealt with {\em
ranked} trees \cite{tata,thomas-handbook}: in such trees, all nodes
have the same fixed number of children (or, a bit more generally, the
number of children of a node is determined by the label of that
node). Over the past several years, however, the focus has shifted
towards {\em unranked} trees, in which there are no restrictions on
the number of children a node can have. For example, the left tree in
Figure \ref{fig:example} is a binary tree in which every
non-leaf node has two children. In the second tree in
Figure \ref{fig:example}, however, different nodes have a different
number of children.
Although unranked trees have been 
considered in the 60s and
70s~\cite{PairQ68,Takahashi75,thatcher67}, and are related to feature trees over
an infinite set of features \cite{smolka92} which are a particular
kind of feature structures that have been investigated by
computational linguists \cite{blackburn,carpenter,rounds86}, their
systematic study was initiated by the development of XML (eXtensible
Markup Language).
XML is a data
format which has become the lingua franca for information exchange on
the World Wide Web. In particular, XML data is typically 
modeled as labeled unranked trees
\cite{neven-csl,vianu-pods}. 

This connection has led to a renewed interest 
in  logical and procedural formalisms
for unranked trees. Since XML trees are used to exchange {\em data},
the usual database query language paradigms apply: one uses logical
formalisms for expressing declarative queries, and procedural
formalisms for evaluating those declarative queries. 
Logics over unranked trees defining a variety of query languages for
them appeared in large numbers over the past 7--8 years, and they come
in many flavors and shapes.
What is common to them, however, is a close connection to  automata models,
and quite often to temporal and modal logics, especially when one describes
properties of paths through a document.

Let us now review some of the parameters according to which logics for
unranked trees can be classified.

\begin{description}
\item[The yardstick logic] Most formalisms are ``inspired'' by one of
the two logics often used in the context of trees: {\em first-order logic}
(\FO), and {\em monadic second-order logic} (\MSO) that extends \FO\
by quantification over sets of nodes. Query languages and schema
formalisms for XML tend to use \MSO\ as the yardstick: for example,
XML Document Type Definition (DTDs, or, more precisely, XSD -- XML
Schema Definition) are essentially equivalent to
\MSO\ sentences, and various languages for extraction of data from XML
documents, although being syntactically very different, have the power
of \MSO\ unary queries.
On the other hand, navigational aspects of XML, in particular, logics
capturing various fragments of XPath, are usually closely related to
\FO\ and its fragments.

%Furthermore, we shall aslo see a model-theoretic approach, in which we
%consider an infinite first-order structure whose universe is the set
%of all unranked trees, and obtain some well-known classes of trees by
%studying first-order definability (in the classic model-theoretic
%sense) over that structure.

\item[Arity of queries] Most commonly one considers Boolean or unary
queries. Boolean queries are logical sentences and thus evaluate to
{\em true} or {\em false}. For example, checking if an XML document
conforms to a schema specification is represented by a Boolean query.
 Unary queries correspond to formulae in one
free variable, and thus produce a set of nodes. For example,
extracting sets of nodes, or evaluating XPath expressions relative to
the root naturally give rise to unary queries. 

\item[Complexity of model-checking/query-evaluation]
The model-checking problem asks whether a tree $T$ satisfies a logical
sentence $\phi$, written $T\models\phi$. If $\phi$ is 
an \MSO\
sentence $\phi$, it can be evaluated in linear time 
in the size of $T$, by converting $\phi$ to a tree automaton.
But there  is a price to pay: in terms of
the size of $\phi$, the complexity becomes non-elementary. This type
of trade-off is one of the central issues in dealing with logics over
trees. Similar issues arise with evaluating formulae $\phi(\bar x)$ in
trees, that is, finding tuples $\bar s$ of nodes such that
$T\models\phi(\bar s)$.

\item[Ordered vs.\ unordered trees]
In the standard definition of unranked trees in the XML context,
children of the same node are ordered by a {\em sibling ordering}. If
such an order is present, we speak of ordered unranked trees. In many
cases, however, this ordering is irrelevant, and some unranked tree
models, such as feature trees, do not impose any ordering on
siblings. There is considerable difference between the expressiveness
of logics and automata models depending on the availability of sibling
ordering. The presence of ordering also affects the yardstick logic,
since without order often counting is needed to match the power of
automata models \cite{courcelle-one}.
\end{description} 

The paper is organized as follows. After we give basic definitions in
Section \ref{notations-sec}, we move to logics for ordered trees. In
Section \ref{ordered-sec} we deal with \MSO-related logics,
including syntactic restrictions of \MSO, a datalog-based logic, and the
$\mu$-calculus. In Section \ref{ordered-fo-sec} we turn to \FO-related logics,  present
analogs of LTL and $\ctls$ that have been studied for expressing
navigational properties, and also look at conjunctive queries over
trees. In Section \ref{unordered-sec} we turn to trees that lack the
sibling ordering, and show that in many logics some form of counting
needs to be added to compensate for the missing ordering. We also
review ambient and feature logics over edge-labeled trees. In Section
\ref{automatic-sec}  we look at the model-theoretic approach.
We consider an infinite first-order structure whose universe is the set
of all unranked trees and obtain some well-known classes of trees by
studying first-order definability (in the classic model-theoretic
sense) over that structure.

\section{Trees, logics, and automata}
\label{notations-sec}

\subsection{{Tree domains, trees, and operations on trees}}
Nodes in unranked trees are elements of $\nn^*$ -- that is, finite
strings whose letters are natural numbers. A string $s=n_0n_1\ldots$
defines a path from the root to a give node: one goes to the $n_0$th
child of the root, then to the $n_1$th child of that element, etc. 
We shall write $s_1\cdot s_2$ for the concatenation of strings $s_1$
and $s_2$, and $\e$ for the empty string. 

\newcommand{\ch}{\prec_{{\rm ch}}}
\newcommand{\sbl}{\prec_{{\rm ns}}}
\newcommand{\fc}{\prec_{{\rm fc}}}
\newcommand{\desc}{\ch^*}
\newcommand{\sib}{\sbl^*}

\begin{figure*}
\begin{center}
\setlength{\unitlength}{0.00072in}
\begingroup\makeatletter\ifx\SetFigFont\undefined%
\gdef\SetFigFont#1#2#3#4#5{%
  \reset@font\fontsize{#1}{#2pt}%
  \fontfamily{#3}\fontseries{#4}\fontshape{#5}%
  \selectfont}%
\fi\endgroup%
{\renewcommand{\dashlinestretch}{30}
\begin{picture}(6478,2110)(0,-10)
\path(3450,1933)(1050,1333)
\blacken\path(1159.141,1391.209)(1050.000,1333.000)(1173.693,1333.000)(1159.141,1391.209)
\path(3450,1933)(1950,1333)
\blacken\path(2050.275,1405.421)(1950.000,1333.000)(2072.559,1349.713)(2050.275,1405.421)
\path(3450,1933)(2850,1333)
\blacken\path(2913.640,1439.066)(2850.000,1333.000)(2956.066,1396.640)(2913.640,1439.066)
\path(3450,1933)(3900,1333)
\blacken\path(3804.000,1411.000)(3900.000,1333.000)(3852.000,1447.000)(3804.000,1411.000)
\path(3450,1933)(4650,1333)
\blacken\path(4529.252,1359.833)(4650.000,1333.000)(4556.085,1413.498)(4529.252,1359.833)
\path(3450,1933)(5550,1333)
\blacken\path(5426.375,1337.121)(5550.000,1333.000)(5442.859,1394.812)(5426.375,1337.121)
\dashline{60.000}(1050,1333)(1950,1333)
\path(1830.000,1303.000)(1950.000,1333.000)(1830.000,1363.000)
\dashline{60.000}(1950,1333)(2850,1333)
\path(2730.000,1303.000)(2850.000,1333.000)(2730.000,1363.000)
\dashline{60.000}(2850,1333)(3900,1333)
\path(3780.000,1303.000)(3900.000,1333.000)(3780.000,1363.000)
\dashline{60.000}(3900,1333)(4650,1333)
\path(4530.000,1303.000)(4650.000,1333.000)(4530.000,1363.000)
\dashline{60.000}(4650,1333)(5550,1333)
\path(5430.000,1303.000)(5550.000,1333.000)(5430.000,1363.000)
\path(6000,2083)(5999,2081)(5996,2078)
	(5991,2072)(5984,2062)(5974,2050)
	(5961,2034)(5945,2016)(5928,1996)
	(5909,1974)(5887,1952)(5865,1929)
	(5841,1907)(5817,1886)(5790,1866)
	(5762,1847)(5733,1830)(5701,1815)
	(5667,1802)(5630,1792)(5591,1785)
	(5550,1783)(5510,1785)(5472,1792)
	(5438,1802)(5409,1814)(5386,1829)
	(5367,1845)(5353,1862)(5343,1879)
	(5335,1897)(5330,1915)(5325,1933)
	(5320,1951)(5315,1969)(5307,1987)
	(5297,2004)(5283,2021)(5264,2037)
	(5241,2052)(5212,2064)(5178,2074)
	(5140,2081)(5100,2083)(5059,2081)
	(5020,2074)(4983,2064)(4949,2051)
	(4917,2036)(4888,2019)(4860,2000)
	(4833,1980)(4809,1959)(4785,1937)
	(4762,1914)(4741,1892)(4722,1870)
	(4705,1850)(4689,1832)(4676,1816)
	(4666,1804)(4650,1783)
\blacken\path(4698.862,1896.633)(4650.000,1783.000)(4746.588,1860.271)(4700.908,1849.816)(4698.862,1896.633)
\path(300,2083)(301,2081)(304,2078)
	(309,2072)(316,2062)(326,2050)
	(339,2034)(355,2016)(372,1996)
	(391,1974)(413,1952)(435,1929)
	(459,1907)(483,1886)(510,1866)
	(538,1847)(567,1830)(599,1815)
	(633,1802)(670,1792)(709,1785)
	(750,1783)(790,1785)(828,1792)
	(862,1802)(891,1814)(914,1829)
	(933,1845)(947,1862)(957,1879)
	(965,1897)(970,1915)(975,1933)
	(980,1951)(985,1969)(993,1987)
	(1003,2004)(1017,2021)(1036,2037)
	(1059,2052)(1088,2064)(1122,2074)
	(1160,2081)(1200,2083)(1241,2081)
	(1280,2074)(1317,2064)(1351,2051)
	(1383,2036)(1412,2019)(1440,2000)
	(1467,1980)(1491,1959)(1515,1937)
	(1538,1914)(1559,1892)(1578,1870)
	(1595,1850)(1611,1832)(1624,1816)
	(1634,1804)(1650,1783)
\blacken\path(1553.412,1860.271)(1650.000,1783.000)(1601.138,1896.633)(1599.092,1849.816)(1553.412,1860.271)
\path(3600,283)(3598,284)(3592,286)
	(3583,289)(3569,294)(3550,301)
	(3526,309)(3499,319)(3469,331)
	(3436,344)(3403,358)(3370,373)
	(3337,389)(3305,405)(3275,423)
	(3246,441)(3221,461)(3198,482)
	(3178,505)(3163,530)(3154,556)
	(3150,583)(3154,610)(3163,635)
	(3178,658)(3196,677)(3217,693)
	(3239,705)(3263,714)(3287,721)
	(3312,726)(3337,730)(3363,733)
	(3388,736)(3413,740)(3439,745)
	(3464,752)(3489,761)(3513,773)
	(3536,789)(3557,808)(3576,831)
	(3590,856)(3600,883)(3604,910)
	(3602,936)(3596,961)(3585,984)
	(3572,1005)(3555,1025)(3536,1043)
	(3515,1061)(3493,1077)(3470,1093)
	(3445,1108)(3421,1122)(3397,1135)
	(3375,1147)(3355,1157)(3337,1165)
	(3323,1172)(3300,1183)
\blacken\path(3421.200,1158.289)(3300.000,1183.000)(3395.312,1104.161)(3375.779,1146.758)(3421.200,1158.289)
\put(5925,1708){\makebox(0,0)[lb]{\smash{{{\SetFigFont{10}{12.0}{\rmdefault}{\mddefault}{\updefault}\Blue{child $\ch$}}}}}}
\put(3075,58){\makebox(0,0)[lb]{\smash{{{\SetFigFont{10}{12.0}{\rmdefault}{\mddefault}{\updefault}\Blue{next-sibling $\sbl$}}}}}}
\put(0,1558){\makebox(0,0)[lb]{\smash{{{\SetFigFont{10}{12.0}{\rmdefault}{\mddefault}{\updefault}\Blue{first
	  child $\fc$}}}}}}
\end{picture}
}
\caption{Basic relations for unranked trees}
\label{basic-rel-fig}
\end{center}
\end{figure*}
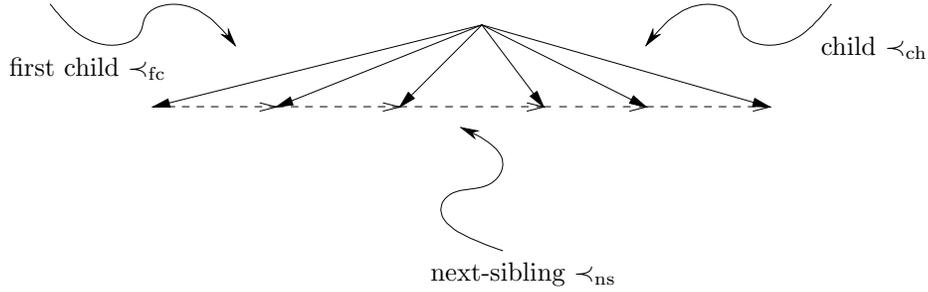

We now define some basic binary relations on $\nn^*$. The {\em child
relation} is 
$$s \ch s' \ \ \LRA \ \ s' = s\cdot i\ \ \text{for some }i \in \nn.$$
The {\em next-sibling} relation is given by:
$$s \sbl s' \ \ \LRA\ \ s=s_0\cdot i\ \ \text{and}\ \ s'=s_0\cdot
(i+1)\ \ \text{for some }s_0\in\nn^*\ \ \text{and }i\in \nn.$$
That is, $s$ and $s'$ are both children of the same $s_0\in\nn^*$, and
$s'$ is next after $s$ in the natural ordering of siblings. We also
use the {\em first child relation}: $s\fc s\cdot 0$. These are shown in Figure \ref{basic-rel-fig}.  

We shall use ${}^*$ to denote the reflexive-transitive closure of a
relation.
%, and ${}^+$ to denote its transitive closure. 
Thus, $\desc$
is the {\em descendant} relation (including self): $s \desc s'$ iff $s$ is a prefix of $s'$ or $s=s'$. The transitive closure of the next-sibling relation,  $\sib$ is a
linear ordering on siblings: $s \cdot i \sib s \cdot j$ iff $i \leq j$. 
We shall be referring to younger/older
siblings with respect to this ordering (the one of the form $s\cdot 0$
is the oldest). 

A set $D \subseteq \nn^*$ is called {\em prefix-closed} if whenever
$s\in D$ and $s'$ is a prefix of $s$ (that is, $s' \desc s$), then
$s'\in D$.

\begin{defi}[Tree domain]
A {\em tree domain} $D$ is a finite prefix-closed subset of $\nn^*$ 
such that
$s \cdot i \in D$ implies $s\cdot j\in D$ for all $j < i$. 
\end{defi}

Let $\Sigma$ be a finite alphabet. We define trees as {\em structures}
that consist of a universe and a number of predicates on the
universe. 

\begin{defi}[$\Sigma$-trees] An {\em ordered unranked tree} $T$
is a structure
$$T=\langle D, \desc, \sib, (P_a)_{a\in \Sigma}\rangle,$$
where $D$ is a tree domain, $\desc$ and $\sib$ are the descendant
relation and the sibling ordering, and the $P_a$'s are interpreted as
disjoint sets whose union is the entire domain $D$.

An {\em unordered} unranked tree is defined as a structure
$\langle D, \desc, (P_a)_{a\in \Sigma}\rangle$, where $D, \desc$, and
$P_a$'s are as above.
\end{defi}

Thus, a tree consists of a tree domain together with a {\em labeling} on its
nodes, which is captured by the $P_a$ predicates: if $s\in P_a$,
then the label of $s$ is $a$. In this case we write $\lambda_T(s)=a$. 

Notice that when dealing with unranked trees we assume that each node
has one label. Later we shall see a connection with temporal logics,
where such a restriction on labeling is normally not imposed. However,
one could always assume unique labeling in that case too, simply by
collecting the set of all labels of a node (in this case the labeling
alphabet becomes $2^\Sigma$).

\subsection{{First-order and monadic second-order logic}}
We shall only consider relational vocabularies, that
is, finite lists $(R_1,\ldots,R_m)$ of relation symbols, each $R_i$ with an
associated arity $n_i$. Over trees, relation symbols
will be either binary (e.g., $\ch, \sbl, \desc$) or unary (the $P_a$'s for
$a\in\Sigma$). 

Formulae of {\em first-order} logic (\FO) are built from atomic
formulae $x=x'$, and $R(\bar x)$, where $x,x'$ are variables, and
$\bar x$ is a tuple of variables whose length equals the arity of $R$,
using the Boolean connectives $\vee, \wedge, \neg$ and quantifiers
$\exists$ and $\forall$. If a formula $\phi$ has free variables $\bar
x$, we shall write $\phi(\bar x)$.
Formulae are evaluated on a structure,
which consists of a universe and interpretations for
relations. Quantifiers $\exists$ and $\forall$ range over the universe
of the structure. For example, an \FO\ formula $$\phi(x)\ \ =\ \ P_a(x) \wedge
\exists y \exists z \big(x \desc y \wedge y \sib z \wedge P_b(y)
\wedge P_c(z)\big)$$ is true for nodes $s$ in a tree $T$ that are
labeled $a$, have a descendant labeled $b$, which in turn has a
younger sibling labeled $c$. 

Formulae of {\em monadic second-order} logic (\MSO) in addition allow
quantification over sets. We shall normally denote sets of nodes by upper case
letters. Thus, \MSO\ formulae have the usual first-order quantifiers
$\exists x\phi$ and $\forall x\phi$ as well as second-order
quantifiers $\exists X\phi$ and $\forall X\phi$, and new atomic
formulae $X(x)$, where $X$ is a second-order variable and $x$ is a
first-order variable. An \MSO\ formula may have both free first-order
and second-order variables. If it only has free first-order
variables, then it  defines a relation on
the universe of the structure.
As an example, an \MSO\ formula 
$\phi_{{\rm odd}}(x,y)$ given by the conjunction of $x \desc y$ and   
$$\exists
X\exists Y
\left(
\begin{array}{cl}
& \forall z\Big(\big(x \desc z \desc y\big)\ \to\ 
\big(X(z)\lra\neg Y(z)\big)\Big) \\
 \wedge & \big(X(x)\wedge Y(y)\big)\\ 
\wedge & \forall z\forall v\ \Big(x \desc z \ch v \desc y \to
\big((X(z)\to Y(v)) \wedge (Y(z)\to X(v))\big)\Big)
\end{array}
\right)$$ says that $y$ is a
descendant of $x$ and the path between them is of odd length. It says that there exist two sets, $X$ and $Y$, 
that partition the path from $x$ to $y$, such that 
$x \in X$, $y \in Y$, and the successor of each 
element in $X$ is in $Y$, and the successor of each element in $Y$ is in $X$. 
In the formula above, $x \desc z \desc y$ is of course an abbreviation
for $(x\desc z)\wedge(z\desc y)$ and likewise for $x \desc z \ch v
\desc y$. 

Note that the relations $\ch$ and $\sbl$ are definable, even in \FO,
from $\desc$ and $\sib$: for example,
$$\neg(x=y)\wedge(x\desc y)\wedge\forall z\big((x\desc
z)\wedge (z\desc y)\to (x=z \vee y=z)\big)$$ defines the child
relation from $\desc$. In \MSO\ one can define $\desc$ from
$\ch$ by stating the existence of a path between two nodes (and
likewise $\sib$ from $\sbl$). However, it is well-known that in \FO\
one {\em cannot} define $\desc$ from $\ch$ (cf.~\cite{FMT}) and this
is why we chose $\desc$ and $\sib$, rather than $\ch$ and $\sbl$, as
our basic relations. However, in all the results about \MSO, we may
assume that the basic relations are $\ch$ and $\sbl$.

In the introduction, we mentioned that we are mostly interested (in
this survey) in Boolean and unary queries. A Boolean query over trees
is just a set of trees closed under isomorphism (that is, a query
cannot distinguish between two isomorphic trees). A unary query $\QQ$ is
a mapping that associates with each tree $T$ a subset $\QQ(T)$ of its
domain. Again, a query is required to be closed under isomorphism. 

\begin{defi}[Definability in logic] Given a logic $\LL$, we say
that a Boolean query (that is, a set $\TT$ of trees) is definable in
$\LL$ if there is a sentence $\phi$ of $\LL$ such that $T \in \TT$ iff
$T \models \phi$. We say that a unary query $\QQ$ 
is definable in $\LL$ if there is a formula $\psi(x)$ of
$\LL$ such that $s \in \QQ (T)$ iff $T\models\psi(s)$, for every tree
$T$ and a node $s$ in $T$.
\end{defi}

\subsection{{Unranked tree automata}}

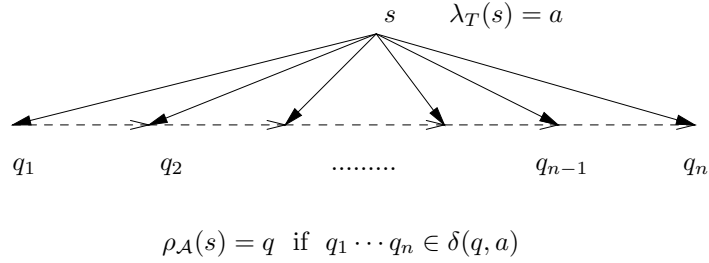
\begin{figure}
\begin{center}
\setlength{\unitlength}{0.0008in}
\begingroup\makeatletter\ifx\SetFigFont\undefined%
\gdef\SetFigFont#1#2#3#4#5{%
  \reset@font\fontsize{#1}{#2pt}%
  \fontfamily{#3}\fontseries{#4}\fontshape{#5}%
  \selectfont}%
\fi\endgroup%
{\renewcommand{\dashlinestretch}{30}
\begin{picture}(4652,1693)(0,-10)
\path(2412,1483)(12,883)
\blacken\path(121.141,941.209)(12.000,883.000)(135.693,883.000)(121.141,941.209)
\path(2412,1483)(912,883)
\blacken\path(1012.275,955.421)(912.000,883.000)(1034.559,899.713)(1012.275,955.421)
\path(2412,1483)(1812,883)
\blacken\path(1875.640,989.066)(1812.000,883.000)(1918.066,946.640)(1875.640,989.066)
\path(2412,1483)(2862,883)
\blacken\path(2766.000,961.000)(2862.000,883.000)(2814.000,997.000)(2766.000,961.000)
\path(2412,1483)(3612,883)
\blacken\path(3491.252,909.833)(3612.000,883.000)(3518.085,963.498)(3491.252,909.833)
\path(2412,1483)(4512,883)
\blacken\path(4388.375,887.121)(4512.000,883.000)(4404.859,944.812)(4388.375,887.121)
\dashline{60.000}(12,883)(912,883)
\path(792.000,853.000)(912.000,883.000)(792.000,913.000)
\dashline{60.000}(912,883)(1812,883)
\path(1692.000,853.000)(1812.000,883.000)(1692.000,913.000)
\dashline{60.000}(1812,883)(2862,883)
\path(2742.000,853.000)(2862.000,883.000)(2742.000,913.000)
\dashline{60.000}(2862,883)(3612,883)
\path(3492.000,853.000)(3612.000,883.000)(3492.000,913.000)
\dashline{60.000}(3612,883)(4512,883)
\path(4392.000,853.000)(4512.000,883.000)(4392.000,913.000)

\put(12,583){\makebox(0,0)[lb]{\smash{{{\SetFigFont{10}{12.0}{\rmdefault}{\mddefault}{\updefault}$q_1$}}}}}
\put(987,583){\makebox(0,0)[lb]{\smash{{{\SetFigFont{10}{12.0}{\rmdefault}{\mddefault}{\updefault}$q_2$}}}}}
\put(2112,583){\makebox(0,0)[lb]{\smash{{{\SetFigFont{10}{12.0}{\rmdefault}{\mddefault}{\updefault}.........}}}}}
\put(3462,583){\makebox(0,0)[lb]{\smash{{{\SetFigFont{10}{12.0}{\rmdefault}{\mddefault}{\updefault}$q_{n-1}$}}}}}
\put(4437,583){\makebox(0,0)[lb]{\smash{{{\SetFigFont{10}{12.0}{\rmdefault}{\mddefault}{\updefault}$q_n$}}}}}

\put(2462,1558){\makebox(0,0)[lb]{\smash{{{\SetFigFont{10}{12.0}{\rmdefault}{\mddefault}{\updefault}$s\
	  \ \ \ \ \ \lambda_T(s)=a$}}}}}

\put(1000,58){\makebox(0,0)[lb]{\smash{{{\SetFigFont{10}{12.0}{\rmdefault}{\mddefault}{\updefault}{$\rho_\AA(s)=q$\ \ if\ \ 
	  $q_1\cdots q_n \in \delta(q,a)$}}}}}}
\end{picture}
}
\caption{Run of an unranked tree automaton}
\label{aut-run-fig}
\end{center}
\end{figure}

A {\em nondeterministic unranked tree automaton, NUTA}
 \cite{thatcher67,bmw},
 over
$\Sigma$-labeled trees is a triple
$\AA=(Q,F,\delta)$ where $Q$ is a finite set of states, $F\subseteq Q$
is the set of final states, and $\delta$ is a mapping $Q \times \Sigma \to
2^{Q^*}$ such that $\delta(q,a)$ is a regular language over $Q$
(normally represented by a regular expression over $Q$). 
A {\em run} of $\AA$ on a tree $T$ with domain $D$ is a function
$\rho_\AA:D \to Q$ such that:
\begin{quote}
if $s$ is a node with $n$ children, and it is labeled $a$, then the
string \\ $\rho_\AA(s\cdot 
0)\cdots \rho_\AA(s \cdot (n-1))$ is in $\delta(\rho_\AA(s),a)$.
\end{quote}
This is illustrated in Figure \ref{aut-run-fig}.
In particular, if $s$ is a leaf labeled $a$, then $\rho_\AA(s)=q$
implies that $\e \in \delta(q,a)$. 
A run is {\em accepting} if $\rho_\AA(\e)\in F$, that is, the root is
in an accepting state. A tree $T$ is {\em accepted} by $\AA$ if there exists
an accepting run. We let $L(\AA)$ denote the set of all trees accepted
by $\AA$. Such sets of trees will be called {\em regular}.

There could be different representations of NUTAs, depending on how
regular expressions over $Q$ are represented. These issues are
discussed in \cite{neven-csl,MN05}.

\subsection{{Binary trees and translations}}
A {\em binary tree domain} is a prefix-closed subset $D$ of
$\{0,1\}^*$ such that if $s \cdot i \in D$, then $s\cdot (1-i)\in D$
(that is, a node is either a leaf, or both its children are in $D$). 
It is common to define (binary) tree automata with both initial and
final states, using the initial states to avoid conditions 
$\e \in \delta(q,a)$ imposed in the runs of NUTAs. That is, a (binary)
{\em nondeterministic tree automaton, NTA}, is a quadruple
$\AA_b=(Q,q_0,F,\delta)$ where $Q$ and $F$ are as before, $q_0$ is the
initial state, and $\delta$ is a function $Q \times Q \times \Sigma
\to 2^Q$. In this case a run $\rho_{\AA_b}$ on a binary tree $T$ with
domain $D$ is a function
from $D$ to $Q$ such that 
if $s$ is a leaf labeled $a$, then
$\rho_{\AA_b}(s)\in\delta(q_0,q_0,a)$, and if 
$s\cdot 0, s\cdot 1$ belong to $D$, and $s$ is labeled $a$, then
$\rho_{\AA_b}(s) \in \delta(\rho_{\AA_b}(s\cdot 0),
\rho_{\AA_b}(s\cdot 1), a)$. As before, a run is accepting if
$\rho_{\AA_b}(\e)\in F$, and $L(\AA_b)$ is the set of all
binary trees for which there exists an accepting run of $\AA_b$.
Sets of trees of this form are regular sets (of binary trees). 

There is a well-known regularity-preserving translation between
unranked and ranked trees. It was first used 
in \cite{rabin} to show decidability of $\SwS$ (but here we shall
apply it only to finite tree domains). The
idea of the translation is that the first successor in the binary tree
corresponds to the first child, and the second successor to the next
sibling. More precisely, we define a mapping $\RT: \nn^*\to\{0,1\}^*$
such that $\RT(\e)=\e$, and if $\RT(s)=s'$, where $s=s_0\cdot i$, then
$\RT(s\cdot 0)=s'\cdot 0$ and $\RT(s_0\cdot (i+1))=s'\cdot 1$. 
Or, equivalently, $\RT(\e)=\e$, and if $\RT(s)=s'$, then $\RT(s\cdot
i)=s'\cdot 0 \cdot 1^i$.

\begin{figure}
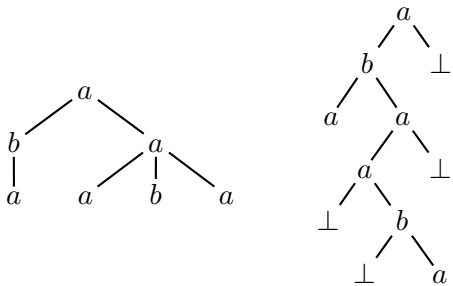
\rm
\begin{center}
\begin{minipage}{4cm}
\pstree{\TR{$a$}}{\pstree{\TR{$b$}}{\TR{$a$}}
\pstree{\TR{$a$}}{\TR{$a$}\TR{$b$}\TR{$a$}}}
\end{minipage}
\begin{minipage}{4cm}
\pstree{\TR{$a$}}{\pstree{\TR{$b$}}
{\TR{$a$}
   \pstree{\TR{$a$}}{
     \pstree{\TR{$a$}}{\TR{$\bot$}   \pstree{\TR{$b$}}{\TR{$\bot$}{\TR{$a$}}}  }
   \TR{$\bot$}}      }
\TR{$\bot$}}
\end{minipage}
\caption{\label{fig:rabin-ex} {\rm A unranked tree $T$ and its
translation $\RT(T)$}}
\end{center}
\end{figure}

If $D$ is
an unranked tree domain, we let $\RT(D)$ be $\{\RT(s)\mid s\in D\}$
together with $\RT(s)\cdot 1$ if $s$ is a non-leaf last child, and
$\RT(s)\cdot 0$ if $s$ a leaf, other than the last sibling (these
additions ensure that $\RT(D)$ is a binary tree domain). We define
$\RT(T)$ to be a tree with domain $\RT(D)$, where $\RT(s)$ has the same
label as $s$, and the added nodes are labeled by a symbol $\bot\not\in
\Sigma$. An example is shown in Figure \ref{fig:rabin-ex}.

The following is a folklore result.

\begin{lemma}
\label{nuta-nta-lemma}
For every NUTA $\AA$, there is an NTA $\AA_b$ such that 
$L(\AA_b)=\{\RT(T)\mid T \in L(\AA)\}$, and conversely, for every
NTA $\AA_b$ there is an NUTA $\AA$ such that the above holds.
\end{lemma}

\newcommand{\dlog}{{\sc Dlogspace}}
\newcommand{\ptime}{{\sc Ptime}}
\newcommand{\np}{{\sc NP}}

Moreover, $\AA_b$ can be constructed from $\AA$ very fast, in \dlog\
\cite{GKS-jacm}.

Other regularity-preserving translations from unranked trees to binary
trees exist. For example, \cite{niehren-rta04} views unranked trees as
built from labeled nodes by means of a binary operation $T@T'$ that
attaches $T'$ at the new youngest child of the root of $T$.
This immediately yields a binary tree representation and an automaton
construction, and of course an analog of Lemma \ref{nuta-nta-lemma}
holds.

\section{Ordered trees: MSO and its relatives}
\label{ordered-sec}

In the next two sections we only deal with ordered unranked trees. 

As we mentioned already, \MSO\ is often used as a yardstick logic for
trees, because of its close connection to regular languages. 
The following result belonged to folklore, and was explicitly stated
in \cite{nevenphd}.

\begin{thm} 
\label{mso-thm}
A set of unranked trees is regular iff it is definable in \MSO.
\end{thm}

When restricted to strings and binary trees, this 
corresponds to well-known results by B\"uchi \cite{buchi} saying that
\MSO\ equals regular languages over strings, and by Thatcher,
Wright \cite{tw}, and Doner \cite{doner}, saying that \MSO\ equals
regular (binary) tree languages.

\newcommand{\SAT}{\text{{\sc Sat}}}

There is also a close connection between automata, \MSO, and the
common formalism for describing schemas for XML documents called
DTDs, which are essentially extended context-free grammars. A DTD $d$
over an alphabet $\Sigma$ is a collection of rules $a 
\to e_a$, where $a\in \Sigma$ and $e_a$ is a regular expression over
$\Sigma$. We shall assume there is at most one such rule for each
$a\in\Sigma$. A $\Sigma$-labeled tree $T$ satisfies $d$, if for each
node $s$ of $T$ with $n$ children, and $\lambda_T(s)=a$, the string
$\lambda_T(s\cdot 0)\cdots \lambda_T(s\cdot (n-1))$ is in the language
denoted by $e_a$. We write $\SAT(d)$ for the set of trees that satisfy $d$. 

Each DTD is easily definable by an unranked tree automaton: in fact
its states just correspond to labels of nodes. This, however, is too
restrictive to capture full definability in \MSO.
In fact, DTDs (that is, sets of the form $\SAT(d)$) are closed under neither 
unions nor complement, which makes DTDs unsuitable for capturing a logic with 
disjunction and negation.

However, a slight
extension of DTDs does capture \MSO. An {\em extended DTD} over
$\Sigma$ is a triple $(\Sigma',d',g)$ where $\Sigma'\supseteq \Sigma$,
with $g$ being a mapping $g: \Sigma'\mapsto\Sigma$, and
$d'$ is a DTD over $\Sigma'$. We say that a $\Sigma$-labeled tree $T$
satisfies  $(\Sigma',d',g)$ if there is a $\Sigma' $-labeled tree $T'$
that satisfies $d'$ such that
$T=g(T')$ (more formally, $T$ is obtained by replacing each label $a$
in $T'$ by $g(a)$). We  write $\text{{\sc Sat}}(\Sigma',d',g)$ for
the set of trees that satisfy  $(\Sigma',d',g)$.

The following was established in \cite{thatcher67} and then restated
using the DTD terminology in \cite{vianu00,vianu-pods}.

\begin{proposition} 
A set of unranked trees is \MSO\ definable iff it is of the form 
$\text{{\sc Sat}}(\Sigma',d',g)$ for some  extended DTD $(\Sigma',d',g)$.
\end{proposition}

\newcommand{\QA}{{{\mathcal Q}\!{\mathcal A}}}
\newcommand{\UA}{{\AA_c}}
\newcommand{\UQA}{{\QA_c}}

Theorem \ref{mso-thm} talks about \MSO\ sentences, but it can be
extended to unary \MSO\ queries using the concept of {\em query
automata} \cite{QA}. A (nondeterministic) {\em query automaton} over
unranked $\Sigma$-labeled trees is a quadruple $\QA=(Q,F,\delta,S)$ where
$\AA=(Q,F,\delta)$ is an UNTA, and $S$ is a subset of $Q \times
\Sigma$. Such a query automaton defines two unary queries 
$\QQ^\exists_{\QA}$ and $\QQ^\forall_{\QA}$
on unranked
trees:
\begin{description}
\item[{\rm {\em Existential semantics query}}]: $s \in \QQ^\exists_{\QA}(T)$
iff $(\rho_\AA(s),\lambda_T(s))\in
S$ for some accepting run $\rho_\AA$.
\item[{\rm {\em Universal semantics query}}]: $s \in \QQ^\forall_{\QA}(T)$
iff  $(\rho_\AA(s),\lambda_T(s))\in
S$ for every accepting run $\rho_\AA$.
\end{description}

\begin{thm}
\label{qa-thm}
{\rm (see \cite{QA,nevenphd,grohe-koch-lics})}\
For a unary query $\QQ$ on unranked trees, the following are equivalent:
\begin{enumerate}\itemsep=0pt
\item $\QQ$ is definable in \MSO;
\item $\QQ$ is of the form $\QQ^\exists_\QA$ for some query automaton
$\QA$;
\item $\QQ$ is of the form $\QQ^\forall_\QA$ for some query automaton
$\QA$.
\end{enumerate}
\end{thm}

Query automata, just as usual tree automata, have a deterministic
counterpart; however, in the deterministic version, two passes over
the tree are required. See \cite{QA} for details.

Theorems \ref{mso-thm} and \ref{qa-thm} are constructive. In
particular,  every \MSO\ sentence $\phi$ can
be effectively transformed into an automaton $\AA_\phi$ that accepts a
tree $T$ iff $T\models \phi$. Since tree automata can be determinized,
this gives us a $O(\|T\|)$ algorithm to check whether $T\models \phi$,
if $\phi$ is fixed.\footnote{We use the notation $\|T\|, \|\phi\|$ to
denote the sizes of natural encodings of trees and formulae.} However,
it is well-known that the size of 
$\AA_\phi$ (even for string automata) cannot be bounded by an
elementary function in $\|\phi\|$ \cite{meyer-stockmeyer}. An even stronger
result of \cite{grohe-lics02} says that there could be no algorithm
for checking whether $T\models\phi$ that runs in time
$O(f(\|\phi\|)\cdot \|T\|)$, where $f$ is an elementary function,
unless \ptime=\np. 

Nonetheless, these results do not rule out the existence of a logic
$\LL$ that has the same power as \MSO\ and yet permits faster
model-checking algorithms. Even looking at a simpler case of \FO\ on
strings, where results of \cite{grohe-lics02} also rule out
$O(f(\|\phi\|)\cdot |s|)$ algorithms for checking if a string $s$
satisfies $\phi$, with $f$ being an elementary function, the logic LTL
(linear-time temporal logic) has the same expressiveness as \FO\
\cite{kamp} and admits a model-checking algorithm with running time
$2^{O(\|\phi\|)}\cdot |s|$. 

\subsection{{Logic ETL}} 
The first logic for unranked trees that has the power of \MSO\ and
model-checking complexity matching that of LTL appeared in
\cite{neven00} and 
was called {ETL}
({\em efficient tree logic}). It was obtained by putting syntactic
restrictions on \MSO\ formulae, and at the same time adding new
constructors for formulae, which are not present in \MSO, but are
\MSO-definable.

The atomic formulae of ETL are the same as for \MSO, except that we
are allowed to use both $\ch$ and $\desc$ and are {\em not} allowed to
use the next-sibling relation $\sib$. The formulae of ETL are then
closed under Boolean combinations, {\em guarded quantification}, and
{\em path formulae}. The rules for guarded quantification are as
follows:
\begin{itemize}\itemsep=0pt
\item if $\phi(x,y,X)$ is an ETL formula, then $\exists y\ (x\ch y
\wedge \phi)$ and $\exists y\ (x\desc y
\wedge \phi)$ are ETL formulae; 
\item if $\phi(x,X)$ is an ETL formula, then $\exists X\ (x\desc X
\wedge \phi)$ is an ETL formula. Here $x \desc X$ means that $X$ only
contains descendants of $x$. In this case $\phi$ cannot contain
vertical path formulae (defined below).
\end{itemize}
Path formulae are defined below, and illustrated in Figure \ref{etl-fig}. 
\begin{itemize}\itemsep=0pt
\item if $e$ is a regular expression over ETL  formulae of the form $\psi(u,v)$,
then $e^{\downarrow}(x,y)$ is a (vertical path) ETL formula. The semantics is as
follows: $T\models e^{\downarrow}(s,s')$ if there is a child-relation path 
$s=s_0,s_1,\ldots,s_{n}=s'$ in $T$ and a sequence of
ETL formulae $\psi_i(u,v)$, $i \leq n-1$, such that
$T\models\psi_i(s_i,s_{i+1})$
for each $i \leq n-1$, and the sequence $\psi_0\ldots \psi_{n-1}$
matches $e$. 
\item if $e$ is a regular expression over ETL  formulae of the form $\psi(u,\bar X)$,
then $e^{\rightarrow}(x,\bar X)$ is a (horizontal path) ETL formula.
Then $T \models e^{\rightarrow}(s,\bar X)$ if children $s\cdot i,
i \leq k$ of $s$ can be labeled with ETL formulae
$\psi_i(u,\bar X)$ such that $T\models\psi_i(s\cdot i,\bar X)$ for all
$i$, and the sequence $\psi_0\ldots\psi_k$ matches $e$.
\end{itemize}

\begin{figure}
\begin{center}
\setlength{\unitlength}{0.0008in}
\begingroup\makeatletter\ifx\SetFigFont\undefined%
\gdef\SetFigFont#1#2#3#4#5{%
  \reset@font\fontsize{#1}{#2pt}%
  \fontfamily{#3}\fontseries{#4}\fontshape{#5}%
  \selectfont}%
\fi\endgroup%
{\renewcommand{\dashlinestretch}{30}
\begin{picture}(5337,3703)(0,-10)
\path(2925,3676)(525,676)(5325,676)(2925,3676)
\path(3675,1876)(2775,1576)
\path(3675,1876)(3075,1576)
\path(3675,1876)(3375,1576)
\path(3675,1876)(3675,1576)
\path(3675,1876)(4275,1576)
\path(2625,2926)(2475,2626)(2175,2326)(2325,2026)
\path(1875,1576)(1725,1276)(1875,976)
\put(3825,1576){\makebox(0,0)[lb]{\smash{{{\SetFigFont{14}{16.8}{\rmdefault}{\mddefault}{\updefault}...}}}}}
\put(2250,1876){\makebox(0,0)[lb]{\smash{{{\SetFigFont{34}{40.8}{\rmdefault}{\mddefault}{\updefault}.}}}}}
\put(2175,1726){\makebox(0,0)[lb]{\smash{{{\SetFigFont{34}{40.8}{\rmdefault}{\mddefault}{\updefault}.}}}}}
\put(2025,1651){\makebox(0,0)[lb]{\smash{{{\SetFigFont{34}{40.8}{\rmdefault}{\mddefault}{\updefault}.}}}}}
\put(1875,1576){\makebox(0,0)[lb]{\smash{{{\SetFigFont{34}{40.8}{\rmdefault}{\mddefault}{\updefault}.}}}}}
\put(2700,2851){\makebox(0,0)[lb]{\smash{{{\SetFigFont{10}{12.0}{\rmdefault}{\mddefault}{\updefault}$\phi_0$}}}}}
\put(2550,2551){\makebox(0,0)[lb]{\smash{{{\SetFigFont{10}{12.0}{\rmdefault}{\mddefault}{\updefault}$\phi_1$}}}}}
\put(1950,2251){\makebox(0,0)[lb]{\smash{{{\SetFigFont{10}{12.0}{\rmdefault}{\mddefault}{\updefault}$\phi_2$}}}}}
\put(2400,1951){\makebox(0,0)[lb]{\smash{{{\SetFigFont{10}{12.0}{\rmdefault}{\mddefault}{\updefault}$\phi_3$}}}}}
\put(1300,1201){\makebox(0,0)[lb]{\smash{{{\SetFigFont{10}{12.0}{\rmdefault}{\mddefault}{\updefault}$\phi_{n-1}$}}}}}
%\put(1600,826){\makebox(0,0)[lb]{\smash{{{\SetFigFont{10}{12.0}{\rmdefault}{\mddefault}{\updefault}$\phi_n$}}}}}
\put(2740,1351){\makebox(0,0)[lb]{\smash{{{\SetFigFont{10}{12.0}{\rmdefault}{\mddefault}{\updefault}$\psi_1$}}}}}
\put(3075,1351){\makebox(0,0)[lb]{\smash{{{\SetFigFont{10}{12.0}{\rmdefault}{\mddefault}{\updefault}$\psi_2$}}}}}
\put(3300,1351){\makebox(0,0)[lb]{\smash{{{\SetFigFont{10}{12.0}{\rmdefault}{\mddefault}{\updefault}$\psi_3$}}}}}
\put(3675,1351){\makebox(0,0)[lb]{\smash{{{\SetFigFont{10}{12.0}{\rmdefault}{\mddefault}{\updefault}$\psi_4$}}}}}
\put(4200,1351){\makebox(0,0)[lb]{\smash{{{\SetFigFont{10}{12.0}{\rmdefault}{\mddefault}{\updefault}$\psi_n$}}}}}
\put(2475,2851){\makebox(0,0)[lb]{\smash{{{\SetFigFont{10}{12.0}{\rmdefault}{\mddefault}{\updefault}$x$}}}}}
\put(1975,950){\makebox(0,0)[lb]{\smash{{{\SetFigFont{10}{12.0}{\rmdefault}{\mddefault}{\updefault}$y$}}}}}
\put(3675,1951){\makebox(0,0)[lb]{\smash{{{\SetFigFont{10}{12.0}{\rmdefault}{\mddefault}{\updefault}$x$}}}}}
\put(500,76){\makebox(0,0)[lb]{\smash{{{\SetFigFont{10}{12.0}{\rmdefault}{\mddefault}{\updefault}$e^{\downarrow}({x},{y}):
	  \ \ \ \ 
	  {\phi_0\cdots \phi_{n-1} \in  e}$}}}}}
\put(3525,76){\makebox(0,0)[lb]{\smash{{{\SetFigFont{10}{12.0}{\rmdefault}{\mddefault}{\updefault}${e^{\rightarrow}}({x}):
	  \ \ {\psi_1 \cdots \psi_n \in e}$}}}}}
\end{picture}
}
\caption{The semantics of path formulae of ETL}
\label{etl-fig}
\end{center}
\end{figure}

We also define a slight syntactic modification ${\rm ETL}^\circ$ of
ETL, in which the closure under Boolean connectives is replaced by a
rule that formulae are closed under taking Boolean combinations which
are in DNF: that is, if $\phi_{ij}$'s are ${\rm ETL}^\circ$ formulae,
then $\bigvee_i \bigwedge_j \phi_{ij}'$ is an ${\rm ETL}^\circ$ formula, where
each $\phi_{ij}'$ is either $\phi_{ij}$ or $\neg\phi_{ij}$. Clearly
the expressiveness of ${\rm ETL}^\circ$ is exactly the same as the
expressiveness of ETL.

\begin{thm}
\label{etl-thm}
{\rm (see \cite{neven00})}\ 
With respect to Boolean and unary queries, {\rm ETL} and \MSO\ are
equally expressive. Furthermore, each ${\rm ETL}^\circ$  formula $\phi$ can be
evaluated on a tree $T$ in time $2^{O(\|\phi\|)}\cdot \|T\|$.
\end{thm}

ETL formulae can thus be evaluated in linear time in the size of the
tree, and double exponential time in $\|\phi\|$, by converting Boolean
combinations into DNF. It is not known if ETL itself admits a
$2^{O(\|\phi\|)}\cdot \|T\|$ model-checking algorithm.

\subsection{{Monadic datalog}}
Another approach to obtaining the full power of \MSO\ while keeping
the complexity low is based on the database query language {\em datalog}
(cf.~\cite{AHV}); it was proposed in \cite{GK-jacm,GK-lics}. A datalog
program can be viewed as a prolog program without function
symbols. Datalog is often used to extend expressiveness of database
queries beyond \FO. 

\newcommand{\ggets}{{\mbox{:--}}}

A datalog program consists of a sequence of rules $$H\ \  \ggets\ \ 
P_1,\ldots,P_k,$$ where $H$ and all $P_i$'s are atoms: that is, atomic
formulae of the form $E(\bar x)$. The predicate $H$ is called
the head of the rule, and $P_1,\ldots,P_k$ are called its body. Every
variable that appears in the head is required to appear in the
body. Given a datalog program $\PP$, predicates which appear as a head
of some rule are called intensional, and other predicates are called
extensional. If all intensional predicates are monadic (that is, of
the form $H(x)$), then $\PP$ is a {\em monadic} datalog program. 

Given a datalog program $\PP$ with extensional predicates
$P_1,\ldots,P_m$ and intensional predicates $H_1,\ldots,H_\ell$, and a
structure $\DD=\langle D,P^\DD_1,\ldots,P^\DD_m\rangle$ 
that interprets each $p$-ary predicate $P_i$ as $P_i^\DD \subseteq 
D^p$, we define $\PP(\DD)$ as the least fixed point of the {\em
immediate consequence} operator. This operator takes a structure
$\HH'=\langle D,H_1',\ldots,H_\ell'\rangle$ and produces
another structure $\HH''=\langle
D,H_1'',\ldots,H_\ell''\rangle$ such
that a tuple $\bar a$ is in $H_i''$ if it is in $H_i'$ or there is a
rule $H_i(\bar x) \ggets R_1(\bar x,\bar y),\ldots,R_s(\bar x,\bar y)$
and a tuple $\bar b$ such that for each extensional predicate $R_i$,
the fact $R_i(\bar a,\bar b)$ is true in $\DD$, and for each intensional
predicate $R_i$, the fact $R_i(\bar a,\bar b)$ is true in $\HH'$.

A monadic datalog query is a pair $(\PP,H)$ where $\PP$ is a monadic
datalog program, and $H$ is an intensional predicate. The value of $H$
in $\PP(\DD)$ is the output of this program on $\DD$.

\newcommand{\leaf}{\text{\sl Leaf}}
\newcommand{\lastch}{\text{\sl LastChild}}
\newcommand{\rt}{\text{\sl Root}}

We consider three unary predicates on unranked tree domains: $\leaf$,
$\lastch$, and $\rt$. Given a tree domain $D$, they are interpreted as
$$
\begin{array}{rcl}
\leaf & = & \{s\in D\mid \neg \exists s'\in D: s \ch s'\},\\ 
\lastch & = & \{s \cdot i \in D\mid s \cdot (i+1)\not\in D\},\\
\rt & = &\{\e\}.
\end{array}
$$ 

\begin{thm}
\label{gk-thm}
{\rm (see \cite{GK-jacm})}\ 
A unary query over unranked trees is definable in \MSO\ iff it is
definable in monadic datalog over extensional predicates $\fc$,
$\sbl$, $\leaf$, $\lastch$, $\rt$, and $P_a, a \in \Sigma$.

Furthermore, each monadic datalog query $(\PP,H)$ can be evaluated on
a tree $T$ in time $O(\|\PP\|\cdot \|T\|)$. 
\end{thm}

There are two proofs of this result in \cite{GK-jacm}: one codes query
automata in monadic datalog, and the other one uses the standard
reduction to ranked trees and the composition method
(cf.~\cite{thomas-ctl}) for \MSO\ games.

\subsection{{$\mu$-calculus}}
Yet another way of getting a logic equivalent to \MSO\ is suggested by
a close connection between \MSO\ and the modal $\mu$-calculus $\muc$
on ranked trees, which can easily be extended to the unranked case by
using the connection between ranked and unranked trees.
 It was shown in \cite{emerson-focs91,niwinski88} that
every property 
of  infinite binary trees definable in \MSO\ is also  definable in
$\muc$. To deal with unranked trees, we shall define $\muc$ over
$\Sigma$-labeled structures that have several binary relations
$E_1,\ldots,E_m$, 
cf.~\cite{niwinski-book}. Formulae 
of 
$\calcmu$ are given by 
$$\phi \  \mbox{:=} \ a\ (a\in \Sigma) \mid X\mid \phi\vee\phi\mid
\neg\phi\mid \dm(E_i)\phi\mid 
\mu X\;\phi(X),$$
where in $\mu X\;\phi(X)$, the variable $X$ must occur positively in
$\phi$. Given a structure $T$ with domain $D$, $s\in D$, and a valuation
$v$ for free variables (such that each $v(X)$ is a subset of $D$), 
we define the semantics by
\begin{itemize}\itemsep=0pt     
\item
$(T, v, s) \models a$ iff $s$ is labeled $a$.
\item
$(T, v, s) \models \phi\vee\phi'$ iff $(T, v, s) \models
  \phi$ or  $(T, v, s) \models \phi'$.
\item
$(T, v, s) \models \neg\phi$ iff $(T, v, s) \models \phi$ is false. 
\item
$(T, v, s) \models X$ iff $s \in v(X)$. 
\item
$(T, v, s) \models \dm(E_r) \phi$ iff
$(T,v,s')\models \phi$ for some $s'$ with $(s,s')\in E_r$.
\item
$(T, v, s) \models \mu X \,\phi(X)$ iff $s$ is in the least fixed
  point of the operator defined by $\phi$; in other words, 
 if $$s \in \bigcap \{P
  \mid \{s' \mid (T, v[P/X],s') \models
  \phi\} \subseteq P\},$$ where $v[P/X]$ extends the valuation $v$ by
  $v(X)=P$. 
\end{itemize}
We shall list explicitly binary relations $E_i$, writing
$\calcmu[E_{1},\ldots,E_m]$ to refer $\calcmu$
formulae that only use those relations. 
An $\muc$ formula $\phi$ without free variables
naturally defines a unary query on trees ($\{s \mid
(T,s)\models\phi\}$) and a Boolean query on trees (by checking if
$(T,\e)\models\phi$). 

Using the translation into ranked trees (or direct coding of
automata), it is easy to show the following (see \cite{lics05}):
\begin{proposition}
\label{bl-prop}
The class of Boolean  \MSO\ queries on unranked trees is
precisely the class of Boolean queries defined by
$\muc[\fc,\sbl]$.
\end{proposition}

If we consider unranked trees as structures with relations
$\fc$ and $\sbl$, then they are acyclic, and hence the
complexity of model checking is
$O(\|\phi\|^2\cdot\|T\|)$ \cite{mateescu}. Furthermore, 
results of \cite{mateescu} tell us that one can strengthen Proposition
\ref{bl-prop}: \MSO\ equals alternation-free $\muc$ over $\fc,\sbl$.
For alternation-free $\muc$ formulae over unranked trees the
complexity of model-checking further reduces to
 $O(\|\phi\|\cdot \|T\|)$, matching the complexity of monadic
datalog. 

It is also possible to characterize unary \MSO\ queries over unranked trees
in terms of the {\em full} $\mu$-calculus $\fmuc$
(cf.~\cite{vardi-icalp98}) which adds
backward modalities $\dm(E_i^-)\phi$ 
with the
semantics 
\begin{itemize}
\item $(T,s)\models \dm(E_i^-)\phi$ iff $(T,s')\models \phi$ for
some $s'$ such that $(s',s)\in E_i$.
\end{itemize}

\begin{proposition}
\label{bl-thm}
{\rm (see \cite{lics05})}\ 
The class of unary \MSO\ queries on unranked trees is
precisely the class of queries defined by
$\fmuc[\ch,\sbl]$. 
\end{proposition}

There are other fixed-point constructions that have been shown to
capture the power of automata and \MSO\ over unranked trees; see,
e.g. \cite{seidl98}.

\section{Ordered trees: FO and its relatives}
\label{ordered-fo-sec}

We continue dealing with ordered trees, but now we move to logics
closely related to first-order, as opposed to monadic second-order.

While a lot is known about \FO\ on both finite and infinite strings,
it has not been nearly as extensively studied for trees until very
recently. Recall that over strings -- which we can view as
trees with only unary branching -- \FO\ defines precisely the
star-free languages (cf.~\cite{thomas-handbook}), and over both finite
and infinite strings \FO\ has exactly the power of LTL \cite{kamp}. It can
further be characterized by aperiodicity of the syntactic monoid
(cf.~\cite{straubing}).  

In contrast, the natural analog of star-free expressions over binary
trees captures not \FO\ but \MSO\ \cite{potthoff}. Algebraic characterizations of 
\FO-definable classes of binary trees have been obtained very recently
\cite{luc-stacs,igor-concur04,esik-weil}, with \cite{luc-stacs}
showing that \FO-definability (without the descendant relation) is
decidable for regular tree 
languages. One well-known equivalent  logical 
description of \FO\ on binary trees is Hafer-Thomas's theorem \cite{thomas-ctl}
stating that over finite binary trees, $\FO=\ctls$ ($\ctls$ is a
branching time temporal logic
widely used in verification, cf.~\cite{MC-book}, and it will be
defined shortly). Actually, the result of \cite{thomas-ctl} shows that
$\ctls$ is equivalent to \MSO\ with second-order quantification over
paths only, but over finite trees this fragment of \MSO\ is equivalent
to \FO.

The interest in logics over unranked trees whose power is equal to or
subsumed by that of \FO\ stems from the fact that navigational
features of XPath can be described in \FO. XPath \cite{xpath} is a W3C
standard for describing paths in XML documents. 
For example, an XPath expression 
$$/\!/a[/\!/b]/c$$
produces the $c$-labeled children of $a$-labeled nodes having a
$b$-labeled descendant. Here $/\!/$ denotes descendant, $/$ denotes
child, and $[\;]$ is a node test. The expression above looks for
$a$-nodes (descendants of the root) in which the test $[/\!/b]$ is
true (the existence of a node labeled $b$) and from there it proceeds
to children of such nodes labeled $c$. While this is the syntax one
typically finds in the literature on XPath, here we shall use a
different syntax, highlighting connections with temporal logics.

In this section we shall look for connections between XPath, \FO\
on trees, and temporal logics, which are designed to talk about
properties of paths. 

Logics introduced in the context of studying XPath, and more
generally, navigational properties of XML documents, can be roughly
subdivided into two groups. Firstly, one may try to establish analogs
of Kamp's theorem (stating that $\FO={\rm LTL}$ over strings) for
trees.  Secondly, one can try to extend Hafer-Thomas's theorem (the
equivalence $\FO=\ctls$)
from binary to unranked trees. 

Both ways are possible, and in both cases we get FO completeness
results, stating that some temporal logics have precisely the power of
unary FO queries. 

\subsection{{XPath and temporal logics}}

We start with LTL-like logics.
First, recall the syntax of LTL over alphabet $\Sigma$:
$$\phi, \phi'\ \ \mbox{:=}\ \ a, a \in \Sigma \ \mid\ 
\phi \vee \phi' \ \mid\
\neg \phi\ \mid\
\X\phi\ \mid\ \Y\phi\ \mid\ 
\phi\U\phi'\ \mid\
\phi\S\phi'.$$
Formulae of LTL are interpreted over finite or infinite strings over
$\Sigma$: a formula is evaluated in a position in a string. Given 
a string $s=a_0a_1\ldots$, we  
define the semantics  as follows:
\begin{itemize}\itemsep=0pt
\item $(s,i)\models\ a$ iff $a_i=a$;
\item $(s,i)\models \X\phi$ (``next'' $\phi$)
iff
$(s,i+1)\models \phi$; 
\item $(s,i)\models \Y\phi$ iff
$(s,i-1)\models \phi$; 
\item $(s,i)\models\phi\U\phi'$ ($\phi$ ``until'' $\phi'$) if there exists
$j \geq i$ such that 
$(s,j)\models\phi'$ and $(s,k)\models \phi$ for all $i \leq k < j$;
\item the semantics of the dual $\phi\S\phi$ ($\phi$ ``since'' $\phi'$) is that
there exists $j \leq i$ such that $(s,j)\models\phi'$ and 
$(s,k)\models \phi$ for all $j < k \leq i$. 
\end{itemize}
Note that it is possible to avoid $\X$ and $\Y$ by defining a strict
semantics for $\U$ and $\S$, without requiring $\phi$ to be true in
$(s,i)$. 

\newcommand{\TLtree}{\text{\rm TL}^{\text{\rm tree}}}
\newcommand{\TLCtree}{\text{\rm TL}_{\text{\rm count}}^{\text{\rm tree}}}

\newcommand{\ich}{\text{\rm ch}}
\newcommand{\isbl}{\text{\rm ns}}

We now consider a logic $\TLtree$ ({\em tree temporal logic},
cf.~\cite{marx-tods,schl92}) defined as follows: 
$$\phi, \phi'\ \ \mbox{:=}\ \ a, a \in \Sigma \ \mid\ 
\phi \vee \phi' \ \mid\
\neg \phi\ \mid\
\X_*\phi\ \mid\ \Y_*\phi\ \mid\ 
\phi\U_*\phi'\ \mid\
\phi\S_*\phi',$$
where $*$ is either 'ch' (child) or 'ns' (next sibling). 
We define the semantics with respect to a tree $T$ and a node $s$ in $T$:
\begin{itemize}\itemsep=0pt
\item $(T,s)\models a$ iff $\lambda_T(s)=a$;
\item $(T,s)\models \X_\ich\phi$ if $(T,s\cdot i)\models\phi$ for some
$i$;
\item  $(T,s)\models \Y_\ich\phi$ if $(T,s')\models\phi$ for the node
$s'$ such that $s'\ch s$; 
\item $(T,s)\models \phi\U_\ich\phi'$ if there is a node $s'$ such
that $s\desc s'$, $(T,s')\models\phi'$, and for all $s''\neq s'$
satisfying $s\desc s'' \desc s'$ we have $(T,s'')\models \phi$. 
\end{itemize}
The
semantics of $\S_\ich$ is defined by reversing the order in the
semantics of $\U_\ich$, and the semantics of $\X_\isbl, \Y_\isbl,
\U_\isbl$, and $\S_\isbl$ is the same by replacing the child relation
with the next sibling relation.

$\TLtree$ naturally defines unary queries on trees, and it also defines
Boolean queries: we say that $T\models\phi$ if $(T,\e)\models\phi$.

\begin{thm}
\label{marx-tl-thm}
{\rm (see \cite{marx-tods})}\
A unary or Boolean query over unranked trees is definable in \FO\ iff
it is definable in $\TLtree$.
\end{thm}

In $\ctls$-like logics, there are two kinds of formulae: those
evaluated in nodes of trees, and those evaluated on paths in
trees. This is similar to the situation with XPath, which has filter
expressions evaluated on nodes, and location path expressions, which
are evaluated on paths in XML trees. We shall now present two logics:
$\ctls$ with the past, in the spirit of \cite{KP95}, and a $\ctl$-like
reformulation of XPath, as presented in \cite{marx-tods}. 

\newcommand{\XPC}{\text{\rm CXPath}}
\newcommand{\XPCore}{\text{\rm Core\_XPath}}
\newcommand{\step}{\text{{\tt step}}}

We start with XPath-inspired logics, and present them using a slight
modification of the syntax that keeps all the main XPath constructions
and yet makes the connection with temporal
logics more visible.

The language $\XPC$ \cite{marx-tods} (Conditional
XPath) is defined to
have {\em node formulae} $\alpha$ 
and {\em path formulae} $\beta$ given by:
$$
\begin{array}{rcl}
\alpha, \alpha' & \mbox{:=} & a, a \in \Sigma \ \mid\ \neg \alpha\ \mid\
                     \alpha \vee \alpha' \ \mid\ \E\beta\\
\beta, \beta' & \mbox{:=} & ?\alpha\ \mid\ \step \ \mid\ 
                            (\step/?\alpha)^+\ \mid\ 
                            \beta/\beta'\ \mid\ 
                            \beta \vee \beta'
\end{array}
$$
where $\step$ is one of the following: $\ch$, $\ch^-$, $\sbl$, or
$\sbl^-$. 
Intuitively $\E\beta$ states the existence of a path starting in a
given node and satisfying $\beta$, $?\alpha$ tests if $\alpha$ is true
in the initial node of a path, and $/$ is the composition of paths.

Formally, given a tree $T$, we evaluate each node formula in a node (that is, we
define $(T,s)\models\alpha$), and each path formula in two nodes
(that is, 
$(T,s,s')\models\beta$). The semantics is then as
follows (we omit the rules for Boolean connectives):
\begin{itemize}\itemsep=0pt
\item $(T,s)\models a$ iff $\lambda_T(s)=a$;
\item $(T,s) \models \E\beta$ iff there is $s'$ such that
$(T,s,s')\models\beta$;
\item $(T,s,s') \models ?\alpha$ iff $s=s'$ and $(T,s)\models\alpha$;
\item $(T,s,s') \models \step$ iff $(s,s')\in \step$;
\item $(T,s,s')\models \beta/\beta'$ iff for some $s''$ we have
$(T,s,s'')\models \beta$ and $(T,s'',s')\models\beta'$;
\item $(T,s,s')\models (\step/?\alpha)^+$ if there exists a sequence
of nodes $s=s_0,s_1,\ldots,s_k=s'$, $k > 0$, such that each
$(s_i,s_{i+1})$ is in $\step$, and $(T,s_{i+1})\models\alpha$ for each
$i<k$. 
\end{itemize}

The language $\XPCore$ \cite{GKS-jacm,GK-tods} is obtained by only
allowing $\step^+$ as opposed to $(\step/?\alpha)^+$ in the definition
of path formulae. Notice that since $\step^+=(\step/?\text{\em
true})^+$, where $\text{\em true}=\bigvee_{a\in\Sigma}a$, we have 
$\XPCore \subseteq \XPC$. 

The earlier example of an XPath expression ($/\!/a[/\!/b]/c$) can be
represented  in this syntax by a node formula $c \wedge
\E(\ch^-/?a/\ch^+/?b)$ saying that a node is labeled $c$, and there is
a path that starts by going to its parent, finding $a$ there, and then
going to a descendant of that $a$ and finding a $b$.

$\XPCore$ corresponds to XPath as defined by the W3C \cite{xpath}, while
$\XPC$ represents an addition to XPath proposed by
\cite{marx-tods}. This addition is essentially the ``until'' operator
of temporal logic: for example, to represent the strict version of
until (that is, to say that in the next element of a path  $a
\U b$ holds), one could write 
$\ch/?b \ \vee\ (\ch/?a)^+/\ch/?b$. 

Node formulae of either $\XPC$ or $\XPCore$ naturally define unary
queries on trees. These can be characterized as follows.

\begin{thm}
\label{marx-thm} a) {\rm (see \cite{marx-tods})}\ The node formulae of $\XPC$
have precisely the power of \FO\ unary queries.

b) {\rm (see \cite{marx-tdm})}\ The node formulae of $\XPCore$ have
precisely the power of unary $\FO^2$ queries (that is, $\FO$ with two variables)
in the vocabulary  $\ch, \desc, \sbl, \sib$.
\end{thm}

Part a) of Theorem \ref{marx-thm} can also be extended to formulae in
two free variables, see \cite{marx-tods}.

\subsection{{A $\ctls$-like logic}} 
The logics $\ctl$ (computation tree logic) and $\ctls$ are branching
time temporal logics used in verification of reactive systems. They
are normally defined without past connectives, but here we use the
syntax close to that of \cite{KP95} to make it possible to reason
about the past. In these logics, one also has node (usually called
state) formulae and path formulae, but path formulae are evaluated on
paths, not on arbitrary pairs of nodes. 

We define $\ctlspast$ 
node formulae $\alpha$, 
and child and sibling path formulae $\beta_\ich$ and $\beta_\isbl$, 
as follows:
\begin{eqnarray*}
\alpha, \alpha'
 \ &:=&  \ a \ (a \in \Sigma) \ \mid \ \neg \alpha \ \mid
  \ \alpha \vee \alpha' \ \mid \ \E \beta_\ich\ \mid\ \E\beta_\isbl
\\ 
\beta_\ich, \beta_\ich'  \ &:=&  \ \alpha 
\ \mid \neg \beta_\ich \ \mid \ \beta_\ich \vee
\beta_\ich' \ \mid \ \X_\ich \beta_\ich \ \mid\ \Y_\ich\beta_\ich \ 
\mid \ \beta_\ich \U_\ich \beta_\ich'\ \mid\ \beta_\ich \S_\ich \beta_\ich'\\
\beta_\isbl, \beta_\isbl'  \ &:=&  \ \alpha 
\ \mid \neg \beta_\isbl \ \mid \ \beta_\isbl \vee
\beta_\isbl' \ \mid \ \X_\isbl \beta_\isbl \ \mid\ \Y_\isbl\beta_\isbl \ 
\mid \ \beta_\isbl \U_\isbl \beta_\isbl'\ \mid\ \beta_\isbl \S_\isbl \beta_\isbl'
\end{eqnarray*}
Given a tree, a child-path $\pi_{\ich}$ is a sequence of nodes on a path from
the root to a leaf, and a sibling-path is a sequence $\pi_{\isbl}$ of nodes of the
form $s\cdot 0,\ldots, s\cdot (n-1)$ for a node $s$ with $n$
children. We define the semantics of node formulae with respect to a
node in a tree, and of path formulae with respect to a path and a node
on the path (i.e., we define the notion of
$(T,\pi_*,s)\models\beta_*$, for $*$ being '$\ich$' or '$\isbl$'). 
\begin{itemize}\itemsep=0pt
\item $(T,s)\models\E\beta_*$ if there exists a path $\pi_*$ such that
$s\in\pi_*$ and $(T,\pi_*,s)\models\beta_*$;
\item $(T,\pi_\ich,s)\models \X_\ich\beta$ if $(T,\pi_\ich,s')\models
\beta$, where $s'$ is the child of $s$ on path $\pi_\ich$;
\item $(T,\pi_\ich,s)\models \Y_\ich\beta$ if $(T,\pi_\ich,s')\models
\beta$ where $s'$ is the parent of $s$ on $\pi_\ich$;
\item $(T,\pi_\ich,s)\models \beta \U_\ich\beta'$ if for some $s'\neq
s$ such that $s'\in\pi_\ich$ and $s \desc s'$, we have 
$(T,\pi_\ich,s')\models
\beta'$, and for all $s \desc s'' \desc s', s''\neq s'$, we have
$(T,\pi_\ich,s'')\models
\beta$.
\end{itemize}
The definitions for $\S_\ich$ and for sibling-paths are analogous. 

The following can be seen as an analog of the equivalence $\FO=\ctls$
for finite binary trees
\cite{thomas-ctl}. While the proof the connection between ranked and
unranked tree, the straightforward translation from the binary tree fails
because paths over translations of unranked trees may change direction
between child and sibling-paths arbitrarily many times. 
\begin{thm}
\label{ctlpast-thm}
{\rm (see \cite{lics05})}\ 
A unary or Boolean query over unranked trees is definable in \FO\ iff
it is definable in $\ctlspast$.
\end{thm}

\subsection{{Extensions of \FO\ and regular languages}}
Over strings, \FO\ falls short of all regular languages, as it defines
precisely the star-free ones. However, using arbitrary regular
expressions is often convenient in the context of navigating in XML
documents. 

Given a class $\CC$ of regular expressions, define 
$\FO(\CC)^*$ as an extension of \FO\ with the rules: (i) if $e$ is a
regular expression in $\CC$ over $\FO(\CC)^*$ formulae 
$\psi(u,v)$,
then $e^{\downarrow}(x,y)$ is a formula, and (ii) if 
$e$ is a regular in $\CC$ over $\FO(\CC)^*$   formulae $\psi(u)$,
then $e^{\rightarrow}(x)$ is a formula.
The semantics is the same as for the case of ETL. 
If formulae $\psi$ are restricted to be Boolean combinations of atomic
formulae $P_a$, $a\in \Sigma$, we obtain the logic $\FO(\CC)$.

Let {\rm StarFree} be the class of star-free expressions, and 
{\rm Reg}  the class of all regular expressions.

\begin{thm}
\label{foreg-thm}
{\rm (see \cite{neven00})}\ 
a) $\FO(\text{\rm StarFree})=\FO(\text{\rm StarFree})^*=\FO$.

b) 
$\FO(\text{\rm Reg}) \subsetneq \FO(\text{\rm Reg})^* \subsetneq \MSO$.
\end{thm}

For more on $\FO(\text{\rm Reg})$ and $\FO(\text{\rm Reg})^*$ and
their connections with fragments of \MSO\ such as the path logic
\cite{thomas-path}, see \cite{lics03,neven00}. 

\subsection{{Conjunctive queries over unranked trees}}

Conjunctive queries are a very important class of database queries:
they correspond to the $\exists,\wedge$-fragment of \FO. These are the
same queries that can be expressed by selection, projection, and join
in relational algebra, and thus they form the core of database
queries. Their complexity had been studied extensively. In general,
the complexity of evaluating a conjunctive query $\phi$ over a database
$\DD$ is in \np, in terms of both the size of $\phi$ and the size of
$\DD$. In fact, the problem is \np-hard, and there has been a large
body of work on classifying tractable cases (see, e.g.,
\cite{gottlob-jacm01,GSS}). 

In the case of unranked trees, conjunctive queries are formulae of the
form
$$\phi(\bar x)\ = \ \exists \bar y\ R_1 \wedge \ldots \wedge R_k,$$
where each $R_i$ is either $P_a(z)$ or $z \prec z'$, where $z,z'$ are
variables among $\bar x,\bar y$, and $\prec$ is one of $\ch, \desc$,
$\sbl$, or $\sib$. We write $\text{\rm CQ}(\prec_1,\ldots,\prec_m)$
to denote the class of conjunctive queries over unranked trees in
which only unary predicates $P_a$ and binary predicates among
$\prec_i$ can be used. 

If we restrict ourselves to classes of conjunctive queries that use at
most two binary predicates, then there is a complete classification
for the complexity of query evaluation on unranked trees.

\begin{thm}
\label{GK-conj-thm}
{\rm (see \cite{GK-pods04})} The maximal tractable classes of queries 
$\text{\rm CQ}(\prec_1,\ldots,\prec_m)$,
where all $\prec_i$'s are among $\{\ch,\desc$, $\sbl,\sib\}$, are 
$\text{\rm CQ}(\ch,\sbl,\sib)$ and $\text{\rm CQ}(\desc)$; all others
are \np-hard.
\end{thm}

In fact, \cite{GK-pods04} provided a more general (but rather
technical) criterion for checking when evaluation is in \ptime, and
that condition can be used for other relations present in a
query.

Conjunctive queries can also be used to capture all \FO\ over unranked
tree, even if more than one free variable is used, assuming path
formulae of $\XPC$ can be used as atomic predicates. More precisely,
every \FO\ formula $\phi(\bar x)$ over unranked trees is equivalent to
a union of conjunctive queries whose atomic predicates are
$\beta(x,x')$, where $\beta$ ranges over path formulae of $\XPC$
\cite{marx-tods}.

\section{Unordered trees}
\label{unordered-sec}

In unordered trees, nodes can still have arbitrarily many children,
but the sibling ordering $\sbl$ is no longer available. That is, we
view trees as structures 
$$T=\langle D, \desc, (P_a)_{a\in \Sigma}\rangle,$$
where $D$ is a tree domain, $\desc$ is the descendant
relation, and $P_a$'s define the labels on $D$.
 Logics
considered for unordered unranked trees typically introduce some form
of {\em counting}, see
\cite{lics05,courcelle-one,courcelle-five,zilio-popl,ctlstar-mpl,niehren93,schl92,napster-one,napster-two}. 

A simple explanation for this comes from a modified notion of unranked
tree automata and query automata for unordered unranked trees. A {\em
counting  nondeterministic unranked tree automaton} is a tuple
$\UA=(Q,F,\delta)$, where, as before, $Q$ is a set of states,
and $F\subseteq Q$ is a set of final states. Let $V_Q$ be the set of
variables $\{v_q^k\mid q\in Q, k>0\}$. Then the transition function
$\delta$ maps each pair $(q,a)$, for $q\in Q$ and $a\in\Sigma$, into a
Boolean function over $V_Q$. A {\em run} of $\AA$ on an unordered tree
$T$ with domain $D$ is then a mapping
$\rho_{\UA}: D \to Q$ such that if $\rho_\UA(s)=q$ for a node $s$
labeled $a$, then the value of $\delta(q,a)$ is $1$, where each
variable $v_{q_i}^k$ is set to $1$ if $s$ has at least $k$ children
$s'$ with $\rho_\UA(s')=q_i$, and to $0$ otherwise.
A run is accepting if $\rho_{\UA}(\e)\in F$, and the set of unordered
trees accepted by $\UA$ (that is, trees for which there is an
accepting run) is denoted by $L_u(\UA)$. 

A {\em counting query automaton} $\UQA$ is defined as $(Q,F,\delta,S)$
where $S \subseteq Q\times \Sigma$; it selects nodes $s$ in a run $\rho$ 
where $(\rho_{\UA}(s),\lambda_T(s))\in S$. As before, it can be given
both existential and universal semantics. 

The following appears not to have been stated explicitly, although it
follows easily from results in \cite{nevenphd,QA,napster-one}.

\begin{thm}
\label{unordered-mso-thm}
a) A set of unordered unranked trees is \MSO-definable iff it is of the
form $L_u(\UA)$ for a counting  nondeterministic unranked tree
automaton $\UA$. 

b) A unary query over unordered unranked trees is \MSO-definable iff
it is definable by a counting query automaton $\UQA$ under either
existential or universal semantics.
\end{thm}

\subsection{{\MSO\ and \FO\ over unordered trees}}
Now we look at several alternative characterizations of \MSO\ and \FO\
over unordered unranked trees that exploit the counting connection.

Define the {\em counting $\mu$-calculus} $\cmuc$
(cf.~\cite{janin-lenzi}) 
as an extension of $\muc$ with formulae
$\dm^{\geq k}(E)\phi$. The semantics of $(T,s) \models \dm^{\geq
k}(E)\phi$ is as follows: there exist distinct elements
$s_1,\ldots, s_k$ such that $(s,s_i)\in E$ and $(T,s_i)\models
\phi$ for every $1 \leq i \leq k$. 
The next result follows from 
\cite{igor-mso}, as was noticed in \cite{janin-lenzi}: 

\begin{thm}
\label{mso-child-thm}
Over unordered unranked trees, $\MSO$ and $\cmuc[\ch]$ have precisely
the same power with respect to Boolean queries.
\end{thm}

In fact, it is not hard to show that $\MSO$ can be translated into
alternation-free $\cmuc$, and thus evaluated with complexity
$O(\|T\|\cdot \|\phi\|)$, where $\phi$ is an alternation-free $\cmuc$ formula.

For first-order logic, counting extensions of both the temporal logic
$\TLtree$ and $\ctls$ give us analogs of Kamp's and Hafer-Thomas's
theorems. We define $\TLCtree$ as a version of $\TLtree$ in which only
modalities for the child relation are used, but in addition we have
formulae $\X_\ich^k\phi$, with the semantics that $(T,s)\models
\X_\ich^k\phi$ iff there are at least $k$ children $s'$ of $s$ such
that $(T,s')\models\phi$.

We also extend $\ctls$ with counting. In this counting extension 
$\cctl$, we 
have new state formulae $\EX_{\ich}^{k} \alpha$, where $\alpha$ is a
state formula, with the same semantics as above.

\begin{thm}
\label{fo-count-thm}
{\rm (see \cite{ctlstar-mpl,schl92})}\ \ Over unordered unranked trees,
the classes of Boolean queries expressed in \FO, $\TLCtree$,
and $\cctl$ over binary relation $\ch$, are 
the same.
%\begin{enumerate}\itemsep=0pt
%\item \FO\ over the vocabulary of $\ch$ and $\desc$;
%\item $\TLCtree$;
%\item $\cctl$ over binary relation $\ch$.
%\end{enumerate}
\end{thm}

For unary queries, the equivalence $\FO=\TLCtree$ still holds
\cite{schl92}, and  $\FO$ can be shown to be equivalent to an
extension of $\ctls$ with both counting and the past \cite{lics05,rabinovich}. 

Adding counting does not increase the complexity of model-checking in
temporal logics, which is $2^{O(\|\phi\|)}\cdot \|T\|$,
cf.~\cite{MC-book}. 

Unordered fragments of XPath have also been looked at in the
literature. For example, \cite{wenfei-icdt03} showed that the
restriction of positive (no negation) $\XPCore$ that only uses $\ch$ and $\desc$ is
equivalent to existential positive \FO\ formulae over the vocabulary
that includes both $\ch$ and $\desc$. 

\subsection{{Extensions and more powerful counting}}
Consider now a scenario in which we deal with unordered trees, but in
our formulae we can refer to some arbitrary ordering on siblings:
after all, in any encoding of a tree, siblings will come in some
order. Of course we do not want any particular order to affect 
the truth value, so we want our formulae, even if they use an ordering,
to be independent of a particular ordering that was used.

This is the standard setting of {\em order-invariance}, a very
important concept in finite model theory, cf.~\cite{FMT}. We say that
an \MSO\ sentence $\phi$ over vocabulary including $\desc$ and $\sib$
is {\em $\sbl$-invariant} if for any unordered tree $T$ and any two
expansions $T^{\sbl^1}$ and $T^{\sbl^2}$ with sibling-orderings
$\sbl^1$ and $\sbl^2$ we have $T^{\sbl^1}\models\phi\LRA
T^{\sbl^2}\models\phi$. Any $\sbl$-invariant sentence defines a
Boolean query on unordered trees.

We now define $\CMSO$ \cite{courcelle-one} as an extension of $\MSO$
with {\em modulo 
quantifiers}: for each set variable $X$, and $k > 1$, we have set
new formulae $Q_k(X)$ which are true iff the cardinality of $X$ is
congruent to $0$ modulo $k$.

\begin{thm}
\label{cmso-thm}
{\rm (see \cite{courcelle-five})}\
Over unordered unranked trees, $\sbl$-invariant Boolean queries are
precisely 
the Boolean queries definable in $\CMSO$.
\end{thm}

Further extensions in terms of arithmetic power have been
considered in \cite{napster-one,napster-two}. Recall that Presburger
arithmetic refers to the \FO\ 
theory of the structure $\langle \nn, +\rangle$, and it is known that
this structure admits quantifier elimination in the vocabulary
$(+,<,0,1,(\sim_k)_{k\in\nn})$ where $n\sim_k m$ iff $n-m =
0(\text{\rm mod}\ k)$. We next define {\em Presburger \MSO}, called \PMSO, as
an extension of \MSO\ over unordered trees with the following rule:
if $\phi(\bar x, y, \bar X)$ is a  \PMSO\ formula and $\alpha(\bar
v)$ a Presburger arithmetic formula 
with $|\bar X|=|\bar v|=n$, then $[\phi/\alpha](\bar x, y,\bar X)$
is a \PMSO\ formula. Given valuation $\bar s, s_0, \bar S$ for free
variables, with $\bar S=(S_1,\ldots,S_n)$, let $m_i$ be the number of
children of $s_0$ that belong to $S_i$, that is, the
cardinality of the set $\{s' \mid s_0 \ch s' \text{ and }s' \in S_i\}$. Then
$[\phi/\alpha](\bar s,s_0,\bar S)$ is true iff
$\alpha(m_1,\ldots,m_n)$ is true.

It is easy to see that $\MSO \subsetneq \CMSO \subsetneq \PMSO$ over
unordered trees. Still, \PMSO\ is captured by a decidable automaton
model.

Define Presburger unordered tree automata just as counting
automata except that $\delta$ maps pairs from $Q\times \Sigma$ into Presburger
formulae over $v_q$, for $q \in Q$. We interpret $v_q$ as the number of
children in state $q$, and a transition is enabled if the
corresponding Presburger formula is true in this interpretation.
That is, in a run $\rho$ of such an automaton, if $\rho(s)=q$, the
label of $s$ is $a$ and $\delta(q,a)=\chi(v_{q_1}, \ldots, v_{q_m})$,
then $\chi(n_1,\ldots,n_m)$ is true, where $n_i$ is the number of
children $s'$ of $s$ such that $\rho(s')=q_i$.

\begin{thm}
\label{pmso-thm}
{\rm (see \cite{napster-one})}\
Presburger unordered tree automata and \PMSO\ are
equivalent. Furthermore, both emptiness and universality are
decidable for Presburger unordered tree automata.
\end{thm}

Further extensions with counting have been considered for fixed-point
logics \cite{napster-two} and the $\mu$-calculus with
modulo-quantifiers \cite{lics05}.

\subsection{{Edge-labeled unordered trees}}

While in the early days of tree-based data models there was some
debate as to whether labels should be on edges or nodes, the arrival
of XML seems to have settled that dispute. Nonetheless, there are
several areas where edge-labeled trees play a prominent and role, and
traditionally logical formalisms have been designed for such data. 
First, there are logics for {\em feature trees}, which are a special
case of feature structures used extensively in computational
linguistics \cite{carpenter}. Second, in recent work on spatial
logics, used for describing networks and mobile agents
\cite{cardelli-gordon}, one looks at modal logics over unordered
edge-labeled trees.

In the setting of feature trees, one has an infinite set of features
$\FF$, and in an unordered unranked tree every edge is labeled by an
element $f\in\FF$ such that each node $s$ has at most one outgoing
edge labeled $f$ for each $f\in\FF$. Furthermore, nodes may be labeled
by elements of some alphabet $\Sigma$, as before. It is thus natural
to model feature trees as structures $\langle D, (E_f)_{f\in\FF},
(P_a)_{a\in\Sigma}\rangle$ such that the union of all $E_f$'s forms
the child relation of a tree, and no node has two outgoing
$E_f$-edges.

In the context of computational linguistics, one commonly used logic
for feature trees \cite{blackburn} is the propositional modal logic
that, in the context of feature structures (not necessarily trees), is
also often supplemented with path-equivalence, stating that from a
certain node, one can reach another node following two different
paths. This is the setting of the Kasper-Rounds logic
\cite{rounds86}. Over trees, however, path-equivalence is the same as
equality of paths. A more powerful logic
proposed 
in 
\cite{keller} combined the Kasper-Rounds logic with the
propositional dynamic logic. Its formulae are defined  by
$$\phi, \phi' \ \ \mbox{:=} \ \
a, \ a \in \Sigma\ \mid\ 
\phi\vee\phi'\ \mid\
\neg\phi\ \mid\
\dm(e)\phi\ \mid\ e\approx e',$$
where $e,e'$ are regular expressions over $\FF$. Formulae are
evaluated in nodes of a feature tree $T$. We have
$(T,s_0)\models\dm(e)\phi$ if there is a path $(s_0,s_1)\in E_{f_0},
(s_1,s_2)\in E_{f_1}, \ldots, (s_{n-1},s_n)\in E_{f_{n-1}}$ such that
$(T,s_n)\models\phi$ and $f_0f_1\ldots f_{n-1}$ is a word in the
language denoted by $e$. Furthermore, $(T,s)\models e\approx e'$ if
there is a node $s'$ that can be reached from $s$ by a word in $e$ as
well as a word in $e'$. This semantics is normally considered over
graphs, but over trees this is equivalent to saying that there is a
node reachable by an expression in the language denoted by $e\cap
e'$. That is, $e\approx e'$ is equivalent to $\dm(e\cap e')\text{\em
true}$, and
thus the Kasper-Rounds logic is effectively a reachability logic over
trees.

The reader is referred to \cite{keller} for computational linguistics
applications of this logic. In terms of expressiveness it is clearly
contained in \MSO, and if all expressions $e,e'$ are star-free, then
in \FO\ as well, as long as we have the descendant relation.

Automata for feature trees, based on the algebraic approach to
recognizability \cite{courcelle-one}, were considered in
\cite{niehren93} (which also showed that over flat feature trees the
automaton model coincides with a simple counting logic).

\subsection{{An ambient logic for trees}}

\newcommand{\Lf}{\Lambda}
\newcommand{\Rt}{\triangleright}

Ambient logics are modal logics for trees that have been proposed 
in the context of mobile computation \cite{cardelli-gordon} and later
adapted for tree-represented data \cite{cardelli,cardelli-ghelli}. One
views trees as edge-labeled and defines them by the grammar
$$T, T' \ \ \mbox{:=}\ \ \Lf\ \ \mid\ \
T|T'\ \ \mid\ \ a[T], \ a\in\Sigma,$$
with the equivalences that $|$ is commutative and associative,
and that $T|\Lf \equiv T$. Here $\Lf$ is the empty tree, $|$ is the
parallel composition, and $a[T]$ adds an $a$-labeled edge on top
of $T$. If we extend $\equiv$ to a congruence in the natural way, then
every tree is equivalent to one of the form
$a_1[T_1]|\ldots|a_m[T_m]$, which is viewed as a tree whose root
has $m$ outgoing edges labeled $a_1,\ldots,a_m$, with subtrees
rooted at its children being $T_1,\ldots,T_m$. 
  
There were several similar logics proposed in
\cite{cardelli-tldi,cardelli,cardelli-ghelli,cardelli-gordon,zilio-popl}. Here
we consider the
logic from \cite{cardelli-tldi} whose formulae are given by
$$\phi,\phi' \ \mbox{:=}\ \bot\ \mid\ \Lf\ \mid\  
\phi\wedge\phi'\ \mid\
\neg\phi\ \mid\ 
\phi|\phi'\ \mid\
\phi\Rt\phi'\ \mid\ 
a[\phi]\ \mid\ 
\phi@a,\ \ \ \ a \in \Sigma.$$
The semantics is as follows: 
\begin{itemize}\itemsep=0pt
\item $\bot$ is {\em false}; 
\item $\Lf$ is only true
in a tree equivalent to 
$\Lf$;
\item 
$T \models \phi_1|\phi_2$ iff $T\equiv T_1|T_2$ with
$T_i\models\phi_i$, $i=1,2$; 
\item $T\models \phi\Rt\phi'$ if for every $T'$ with 
$T'\models\phi$ we have $T|T'\models \phi'$; 
\item $T\models a[\phi]$ iff
$T\equiv a[T']$ with $T'\models\phi$;
\item $T\models\phi@a$ iff
$a[T]\models\phi$. 
\end{itemize}

Variations appear in the literature, e.g. with the Kleene star
in \cite{zilio-popl} and recursion in \cite{cardelli-ghelli}. 

The study of ambient logics for trees took a very different path
compared to other logics seen in this survey; in particular, the focus
was on type systems for tree languages and thus on proof systems for
logics, rather than model-checking, its complexity, automata models,
and comparison with other logics. Several lines of work closely
resemble those for node-labeled trees: e.g., \cite{zilio-popl}
introduced Presburger conditions on children, defined an automaton
model, and proved decidability, similarly to
\cite{napster-one,napster-two}. 

However, the ambient logic does not take us outside of the \MSO\
expressiveness: this can be seen by 
going from edge-labeled trees to node-labeled ones. The translation is
simple: the label of each edge $(x,y)$  becomes the label of $y$. The
root will have a special label $\rt$ that cannot occur as a label of
any other node. The only modification in the logic is that now we have
formulae $\Lf_a$ for $a\in\Sigma$, which are true in a singleton-tree
labeled $a$. The resulting logic is easily translated into \MSO. For
example, $\phi|\phi'$ states that the children of the root can be
partitioned into two sets, $X$ and $X'$, such that the subtree that
contains all the $X$-children satisfies $\phi$ and the subtree that
contains all the $X'$-children satisfies $\phi'$. For $\phi\Rt\phi'$,
one can consider $\neg(\phi\Rt\phi')$ saying that there exists a tree
$T'$ such that $T'\models\phi$ and $T|T'\models\neg\phi'$, and use
nondeterministic counting automata to guess this tree $T'$. 

Since moving labels from edges to nodes and back can be defined in
\MSO, we see that the ambient logic is embedded into \MSO. However,
to the best of the author's knowledge, this direction has never been
seriously pursued, and the exact relationship between ambient logics
and other logics described in this survey is still not well
understood.

\section{Automatic structures}
\label{automatic-sec}

In this section we look at a different kind of logics for unranked
trees, using the standard approach of model theory. So far we
represented each tree as a structure and looked at definability over
that structure. Now we want to consider structures whose 
universe is the set of {\em all} trees. Definability over such
structures allows us to describe sets of trees and, more generally,
relations over trees. Choosing the right operations on trees, we shall
find structures where definable sets are precisely the regular
languages. Such structures are very convenient for proving that
certain properties of trees are regular, as it is sometimes easier to
define properties logically than to construct automata for them. 

Let
$\tree(\Sigma)$ be the set of all $\Sigma$-labeled unranked trees. We
consider structures of the form $\MMM=\langle \tree(\Sigma), \Omega\rangle$
where $\Omega$ is a set of relation, constant, and function symbols,
interpreted over $\tree(\Sigma)$.

Let $\Def_n(\MMM)$ be the family of {\em $n$-dimensional definable
sets} over $\MMM$: that is, sets of the form
$$\{\bar T \in \tree(\Sigma)^n\ \mid\ \MMM\models\phi(\bar T)\},$$
where $\phi(x_1,\ldots,x_n)$ is a first-order formula in the
vocabulary $\Omega$. We shall be looking at structures $\MMM$ so that
definable sets would be relations definable in \MSO\ or other
logics. In particular, such relations will be given by automata, and
thus structures $\MMM$ of this kind are called {\em automatic
structures}. 

\newcommand{\MA}{\EuFrak{S}}
\newcommand{\MT}{\EuFrak{T}}
\newcommand{\MC}{\MA_{\text{\rm univ}}}
\newcommand{\MTu}{\MT_{\text{\rm univ}}}
\newcommand{\MTuq}{\MT^{\qq}_{\text{\rm univ}}}

\subsection{{Automatic structures on strings}} 
Before we move to trees, we first survey automatic structures over
strings, cf.~\cite{graedel2000,blss-jacm}. In this case we consider
structures of the form $\langle \Sigma^*,\Omega\rangle$. Our first
example has the following relations in $\Omega$:
\begin{itemize}\itemsep=0pt
\item $\prec$ is a binary relation; $s\prec s'$ is true iff $s$ is a
prefix of $s'$;
\item $L_a$, $a\in \Sigma$,  is a unary relation; $L_a(s)$ is true iff
the last symbol 
of $s$ is $a$;
\item $\el$ is a binary relation; $\el(s,s')$ is true iff $|s|=|s'|$.
\end{itemize}
Let $\MC$ be the structure $\langle \Sigma^*, \prec,
(L_a)_{a\in\Sigma}, \el\rangle$. Then $\MC$ is the {\em universal
automatic structure}: that is, relations $\Def_n(\MC)$ are precisely
the regular relations. Following a standard definition --
see, e.g., \cite{frenchreinvention} -- we say that
a relation $S\subseteq (\Sigma^*)^n$ is
{\em regular} iff there is an automaton $\AA$ over alphabet
$(\Sigma\cup\{\#\})^n$ that accepts precisely the strings $[\bar s]$,
for $\bar s=(s_1,\ldots,s_n)\in S$. The length of $[\bar s]$ is
$\max_i |s_i|$, and the $j$th symbol of $[\bar s]$ is a tuple 
$(\sigma_1,\ldots,\sigma_n)$, where $\sigma_i$ is the $j$th symbol of
$s_i$ if $|s_i|\leq j$, and $\#$ otherwise.

Thus, $\Def_1(\MC)$ contains 
exactly the regular languages over $\Sigma$. Furthermore, the
conversion of formulae over $\MC$ to automata is effective
\cite{graedel2000} and the theory of $\MC$ is decidable. In fact
the theory of every structure that is interpretable in $\MC$ is thus
decidable. 

As an example, consider the structure $\langle \qq,
<\rangle$. Since it is isomorphic to $\langle \{0,1\}^*1, \lex\rangle$,
where $\lex$ is the lexicographic ordering (which is easily definable
in $\MC$), we obtain the well-known decidability of $\langle \qq,
<\rangle$. 

A restriction of $\MC$ that does not have the equal length predicate,
that is, $\MA=\langle \Sigma^*, \prec, (L_a)_{a\in\Sigma}\rangle$ is
known to be strictly weaker that $\MC$ in every dimension: in
particular, $\el$ is not in $\Def_2(\MA)$, and $\Def_1(\MA)$ is
precisely the class of star-free languages \cite{blss-jacm}. 

Notice that both the empty string $\e$ and functions $g_a(s)=s\cdot a$
are definable in $\MA$, and hence another well-known theory
interpretable in $\MA$ and $\MC$  is that of unary term algebras. However, it is
known that for binary term algebras, adding relations like $\prec$
results in undecidable theories \cite{muller-niehren,venkataraman}.
In particular, if we want to keep an analog of the $\prec$-relation
(which is \MSO-definable), we cannot introduce an operation
like the $|$ operation in the ambient logic, and still have a
decidable theory.

\subsection{{Automatic structures on trees}}
To get structures over $\tree(\Sigma)$ that define regular languages
and relations\footnote{The notion of regular relations for trees is
obtained in the same way as for strings}, we find natural analogs of
$\prec$, $L_a$, and $\el$ for trees. For two trees $T_1$ and $T_2$
with domains $D_1$ and $D_2$, we
say that $T_2$ is an {\em extension}  of $T_1$, written $T_1 \preceq
T_2$, if 
$D_1 \subseteq D_2$, and the labeling function of $T_2$ agrees with
the labeling function of $T_1$ on $D_1$. It will actually be more
convenient to work with two extension relations: 
\begin{description}
\item[Extension on the
right $\preceqr$]: For $T_1 \preceqr
T_2$, we require that every $s\in D_2 - D_1$ be of the form $s'\cdot
i$ when $s'\cdot j \in D_1$ for some $j < i$.
\item[Extension down $\preceqd$]:
For $T_1 \preceqd
T_2$, we require that every $s\in D_2 - D_1$ have a prefix $s'$ which
is a leaf of $T_1$. 
\end{description}
Clearly $T_1 \preceq T_2$ iff there is $T'$ such
that
$T_1 \preceqr T'$ and $T' \preceqd T_2$, so in terms of definability
we do not lose anything by using $\preceqr$ and $\preceqd$ instead of
$\preceq$.

We define $L_a$ to be true in a tree $T$ if the rightmost node is
labeled $a$. That is, the node $s\in D$ which is the largest with
respect to $\lex$ is labeled $a$. For the analog of $\el$, recall that
in the standard representation of strings  as first-order structures,
the domain is 
an initial segment of $\nn$, corresponding to the length of the
string. Hence, $\el(s_1,s_2)$ means that if strings are represented as
structures, their domains are the same. We thus introduce a predicate
$\domeq$ such that $T_1 \domeq T_2$ iff $D_1=D_2$ (there $D_i$ is the
domain of $T_i$). 

Now we define analogs of $\MC$ and $\MA$:
$$\begin{array}{rcl}
\MTu & \  =\  & \langle \tree(\Sigma),\ \preceqr,\ \preceqd,\ 
(L_a)_{a\in\Sigma},\ \domeq\rangle \\
\MT &\  =\  & \langle \tree(\Sigma),\  \preceqr,\ \preceqd,\
(L_a)_{a\in\Sigma}\rangle 
\end{array}
$$

\begin{thm}
\label{unr-aut-thm}
{\rm (see \cite{lics03})}\ 
a) For every $n \geq 1$, $\Def_n(\MTu)$ is precisely the class of
regular $n$-ary relations over $\tree(\Sigma)$.

b) $\Def_1(\MT)=\Def_1(\MTu)$ is the class of regular unranked tree
languages, but for every $n > 1$, $\Def_n(\MT)\subsetneq
\Def_n(\MTu)$.
\end{thm}

Notice the difference with the string case, where removing $\el$
(domain equality) resulting in a smaller class of one-dimensional
definable sets: star-free languages. On the other hand, even over
binary trees, the notions of star-free and regular coincide
\cite{potthoff}.

Working with $\MTu$ makes it easy to write rather complicated
properties of tree languages, and then Theorem \ref{unr-aut-thm}
implies that those languages are regular.
For example, if $X \subseteq \tree(\Sigma)$ is regular, then the set
of trees $T$ such that all their extensions can be extended on the
right to a tree in $X$ is regular. Indeed, this is defined as $\phi(T)
= \forall T' \big(T \preceq T' \to \exists T'' (T'\preceqr T'' \wedge
\alpha_X(T''))\big)$, where $\alpha_X$ defines $X$ (by Theorem
\ref{unr-aut-thm}, we know such $\alpha_X$ exists). Then Theorem
\ref{unr-aut-thm} again tells us that $\phi$ defines a regular
language. Furthermore, the conversions from formulae to automata are
effective for both $\MT$ and $\MTu$, which implies decidability of
their theories. 

Other logics over unranked trees can be naturally represented over
these structures. Consider, for example, a restriction of first-order
logic over $\MT$ or $\MTu$ in which all quantification is over {\em
branches}. A branch is a tree $T$ such that the set $\{T' \mid T'
\preceq T\}$ is linearly ordered by $\preceq$. Let $\Def_1^\eta$ be
the class of sets of trees (equivalently, Boolean queries over trees)
definable in this restriction. 

\begin{proposition}
\label{eta-prop}
{\rm (see \cite{lics03})}\ 
$\Def_1^\eta(\MT)$ is precisely the class of \FO-definable Boolean
queries over unranked trees, and 
$\Def_1^\eta(\MTu)$ is the class of Boolean queries definable in a
restriction of \MSO\ in which quantification is allowed only over sets
linearly ordered by $\desc$ or by $\sib$.
\end{proposition}

For more results of this type, see \cite{lics03}.

\subsection{{A different view of unranked trees}}

We conclude by presenting a different view of unranked trees and a
different structure for them that makes it easy to talk about about
their extensions in which new children may be inserted between
existing ones. For example, if we have a tree $T$ with domain
$D=\{\e,0,1\}$, and we want to add more children of the root, they
would have to be added on the right, e.g, we may have an extension
with domain $\{\e,0,1,2,3\}$. But what if we want to add a child on
the left of $0$, and two children between $1$ and $2$? Intuitively, we
need a new tree domain $\{\e,-1,0,\frac{1}{3},\frac{2}{3},1\}$ then. 
We now capture this situation and present a different automatic
structure that makes it easy to derive that certain relations on trees
are regular.

A {\em rational unranked tree domain} is a finite prefix-closed subset of
$\qq^*$. Relation $\desc$ is defined for  rational domains
just as before, and relation $\sib$ is now given by $s\cdot r \sib
s\cdot r'$ iff $r \leq r'$.
Then an unranked tree $T$ over 
a rational unranked tree domain is, as before, a structure 
$T=\langle D, \desc, \sib, (P_a)_{a\in\Sigma}\rangle$. 

Let
$\treeq(\Sigma)$ be the set of all unranked trees with rational 
unranked tree domains. Note that different trees in $\treeq(\Sigma)$
may be isomorphic; we denote this isomorphism relation by
$\cong$. There is a natural one-to-one correspondence between 
$\treeq(\Sigma)/\cong$ and $\tree(\Sigma)$.

We define the extension relation $\preceq$ over
trees in $\treeq(\Sigma)$ as before. A {\em branch}, again, is a tree
$T\in\treeq(\Sigma)$ such that the set $\{T'\mid T' \preceq T\}$ is
linearly ordered by $\preceq$. It follows from the definition of 
rational unranked tree domains that the domain of a branch consists of
all the prefixes of some string $s\in\qq^*$; i.e., it is completely 
determined by $s$, which is its unique leaf. Let $L_a(T)$ be true iff
$T$ is a branch whose leaf is labeled $a$, and let $T_1\lex T_2$ be
true iff $T_1$ and $T_2$ are branches with leaves $s_1$ and $s_2$, and
$s_1 \lex s_2$. We then define the structure
$$\MTuq \ \ = \ \ \langle \treeq(\Sigma),\ \preceq,\ \lex,\ \domeq,\
(L_a)_{a\in\Sigma}\rangle.$$  
In this structure it is much easier to reason about tree extensions
that allow one to insert nodes between existing ones, and not only on
the right or under the leaves. But what about definable sets and
relations over $\MTuq$? It turns out that they are all regular. More
precisely, we can interpret $\MTuq$ in $\MTu$: that is, find a
set $X\in \Def_1(\MTu)$, binary relations $R_1,R_2,R_3\in\Def_2(\MTu)$
and sets $Y_a\in\Def_1(\MTu)$, $a\in\Sigma$, such that $\langle
X,R_1,R_2,R_3,(Y_a)_{a\in\Sigma}\rangle$ is isomorphic to $\MTuq$. That is, we
have:

\begin{proposition}
\label{rational-prop}
The structure $\MTuq$ is interpretable in $\MTu$. Furthermore, there
is a definable subset of the image of $\treeq(\Sigma)$ that contains
exactly one representative of each $\cong$-equivalence class.
\end{proposition}

That is, under the mapping $\iota:
\treeq(\Sigma)/\cong\to\tree(\Sigma)$, definable sets and relations over
$\MTuq$ become precisely the regular tree languages (and relations).
Hence, expressing properties of unranked trees in first-order logic
over  $\MTuq$ allows us to conclude easily that certain tree languages
are regular, and thus \MSO-definable.

\newcommand{\vra}{\mbox{\tt <a>}}
\newcommand{\vrca}{\mbox{\tt </a>}}

\newcommand{\vrb}{\mbox{\tt <b>}}
\newcommand{\vrcb}{\mbox{\tt </b>}}
\newcommand{\vrc}{\mbox{\tt <c>}}
\newcommand{\vrcc}{\mbox{\tt </c>}}
\newcommand{\bra}{\bar{a}}
\newcommand{\brb}{\bar{b}}
\newcommand{\brc}{\bar{c}}

\newcommand{\str}{\text{{\rm str}}}

\section{Other directions and conclusions}

We present here a somewhat random sample of other directions that work
on logics for unranked trees has taken or may take in the future. We
concentrate on streaming applications, and then briefly describe other
directions. 

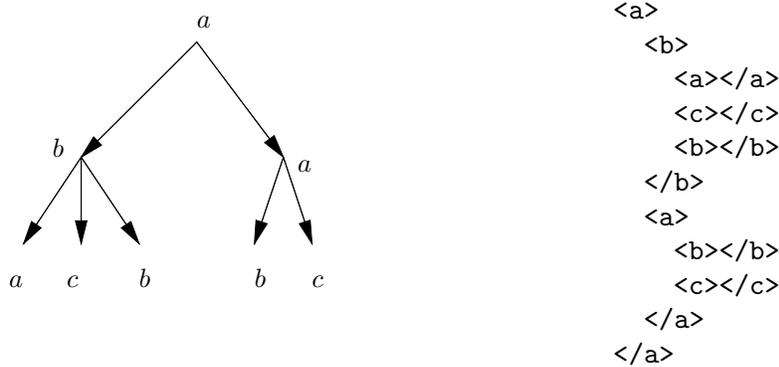
\begin{figure}
\begin{center}
\setlength{\unitlength}{0.001in}
\begingroup\makeatletter\ifx\SetFigFont\undefined%
\gdef\SetFigFont#1#2#3#4#5{%
  \reset@font\fontsize{#1}{#2pt}%
  \fontfamily{#3}\fontseries{#4}\fontshape{#5}%
  \selectfont}%
\fi\endgroup%
{\renewcommand{\dashlinestretch}{30}
\begin{picture}(1698,1500)(0,500)
\path(975,1275)(375,675)
\path(975,1275)(375,675)
\blacken\path(438.640,781.066)(375.000,675.000)(481.066,738.640)(438.640,781.066)
\path(975,1275)(1425,675)
\path(975,1275)(1425,675)
\blacken\path(1329.000,753.000)(1425.000,675.000)(1377.000,789.000)(1329.000,753.000)
\path(375,675)(75,225)
\path(375,675)(75,225)
\blacken\path(116.603,341.487)(75.000,225.000)(166.526,308.205)(116.603,341.487)
\path(375,675)(375,225)
\path(375,675)(375,225)
\blacken\path(345.000,345.000)(375.000,225.000)(405.000,345.000)(345.000,345.000)
\path(375,675)(675,225)
\path(375,675)(675,225)
\blacken\path(583.474,308.205)(675.000,225.000)(633.397,341.487)(583.474,308.205)
\path(1425,675)(1275,225)
\path(1425,675)(1275,225)
\blacken\path(1284.487,348.329)(1275.000,225.000)(1341.408,329.355)(1284.487,348.329)
\path(1425,675)(1575,225)
\path(1425,675)(1575,225)
\blacken\path(1508.592,329.355)(1575.000,225.000)(1565.513,348.329)(1508.592,329.355)
\put(975,1350){\makebox(0,0)[lb]{\smash{{{\SetFigFont{10}{12.0}{\rmdefault}{\mddefault}{\updefault}$a$}}}}}
\put(225,675){\makebox(0,0)[lb]{\smash{{{\SetFigFont{10}{12.0}{\rmdefault}{\mddefault}{\updefault}$b$}}}}}
\put(300,0){\makebox(0,0)[lb]{\smash{{{\SetFigFont{10}{12.0}{\rmdefault}{\mddefault}{\updefault}$c$}}}}}
\put(675,0){\makebox(0,0)[lb]{\smash{{{\SetFigFont{10}{12.0}{\rmdefault}{\mddefault}{\updefault}$b$}}}}}
\put(1500,600){\makebox(0,0)[lb]{\smash{{{\SetFigFont{10}{12.0}{\rmdefault}{\mddefault}{\updefault}$a$}}}}}
\put(1275,0){\makebox(0,0)[lb]{\smash{{{\SetFigFont{10}{12.0}{\rmdefault}{\mddefault}{\updefault}$b$}}}}}
\put(1575,0){\makebox(0,0)[lb]{\smash{{{\SetFigFont{10}{12.0}{\rmdefault}{\mddefault}{\updefault}$c$}}}}}
\put(0,0){\makebox(0,0)[lb]{\smash{{{\SetFigFont{10}{12.0}{\rmdefault}{\mddefault}{\updefault}$a$}}}}}
\end{picture}
} \black
\hspace*{3cm}
\mbox{{
\begin{tabular}{l}
\vra\\
\SEPR\vrb\\
\SEPR\SEPR\vra\vrca\\
\SEPR\SEPR\vrc\vrcc\\
\SEPR\SEPR\vrb\vrcb\\
\SEPR\vrcb\\
\SEPR\vra\\
\SEPR\SEPR\vrb\vrcb\\
\SEPR\SEPR\vrc\vrcc\\
\SEPR\vrca\\
\vrca
\end{tabular}
}}
\caption{An XML document as a tree and as a sequence of tags}
\label{fig-stream}
\end{center}
\end{figure}

\paragraph{Streaming XML documents}
A typical XML document is a sequence of matching opening and closing
tags, with some data between then. For example, the sequence of
opening and closing tags corresponding to a tree is shown in Figure
\ref{fig-stream}. Thus, an XML tree naturally has a string
representation. For example, for the tree in Figure
\ref{fig-stream}, such a representation is
$$aba\bra c\brc b \brb \brb a b \brb c \brc \bra \bra,$$
where we use a label, say $a$, for the opening tag \verb+<a>+, and
$\bra$ for the closing tag
\verb+</a>+.
More generally, for an ordered unranked tree $T$ we define inductively
its string representation $\str(T)$:
\begin{itemize}
\item if $T$ is a single node labeled $a$, then $\str(T)=a\bra$
\item if $T$ has a root labeled $a$, with $n$ children $s_0 \sbl
  \ldots \sbl s_{n-1}$, such that $T_i$ is the subtree rooted at
  $s_i$, $i < n$, then $\str(T)=a\ \str(T_0)\ \ldots\ \str(T_{n-1})\ \bra$.
\end{itemize}

If an XML document $T$ is transmitted as a stream, then the object we work
with is precisely $\str(T)$. Furthermore, we may not have the whole
string $\str(T)$ available, or may need to compute some of its
properties without looking at the whole string (for instance, a device
receiving the stream may have memory limitations and cannot store the
entire stream). One possible model for this scenario was proposed in
\cite{segoufin-pods02}: in this model, one processes the stream
$\str(T)$ by using a finite {\em string} automaton. It is natural to
ask then what kinds of properties of trees can be recognized by finite
automata that run on their streamed representations. More precisely,
one is interested in tree languages of the form
$$L^{\str}_\AA\ = \ \{T \mid \str(T) \text{ is accepted by }\AA\},$$
where $\AA$ is a string automaton.

This question has been primarily addressed in the context of DTD
validation. Namely, given a DTD $d$, is it possible to find an
automaton $\AA_d$ such that
$$L^\str_{\AA_d}\ = \ \SAT(d)?$$
In general, the answer is negative, as was shown in
\cite{segoufin-pods02}. We now sketch a very simple proof of this. 
Consider the following DTD $d_1$:
$$a  \to  ab \ \mid\  ca\ \mid\ \varepsilon,\ \ \ \ \  \ \ 
b  \to  \varepsilon,\ \ \ \ \ \ \ 
c  \to  \varepsilon.
$$
Suppose $\SAT(d_1)=L^\str_{\AA}$ for some $\AA$. The regular language
given by $\AA$ is definable in MSO, say by a sentence of quantifier
rank $r$. Choose numbers $n$ and $k$ so that $a^n$ and $a^{n+k}$
cannot be distinguished by MSO sentences of quantifier rank $r$, and
consider two strings:
$$\begin{array}{rclllll}
s_1 & = & a^n & (ac\brc)^n & a\bra & \bra^n & (b\brb \bra)^n\\
s_2 & = & a^{n+k} & (ac\brc)^n & a\bra & \bra^{n+k} & (b\brb \bra)^n
\end{array}$$
which in turn (by a standard composition argument, see, e.g., \cite{FMT,thomas-handbook}) 
cannot be distinguished by $\AA$. One clearly has $s_1 =
\str(T_1)$ for some $T_1\in\SAT(d_1)$, and $s_2=\str(T_2)$ for a tree
$T_2\in\SAT(d_2)-\SAT(d_1)$, where $d_2$ is 
$$
a  \to  a\ \mid\ ab \ \mid\  ca\ \mid\ \varepsilon,\ \ \ \ \ \ \ 
b  \to  \varepsilon,\ \ \ \ \ \ \ 
c  \to  \varepsilon,
$$ 
which contradicts the assumption $\SAT(d_1)=L^\str_{\AA}$.

While \cite{segoufin-pods02} provides many results on streamed
validation of DTDs, the problem of  characterizing DTDs 
that can be checked by finite automata over streamed
representations remains open. Such a characterization can be found for
\MSO-definable properties as follows. Given an \MSO\ sentence $\phi$
over ordered unranked trees, we say that $\phi$ is streamable if $\{T
\mid T \models\phi\}$ is of the form $L^\str_\AA$ for some finite
string automaton $\AA$. 

\newcommand{\rma}{{\text{{\rm rma}}}}
\newcommand{\rms}{{\text{{\rm rms}}}}

Let $s$ be a node in a tree $T$; define $\rma(s)$ (the right-most
ancestor) to be the smallest prefix of $s$ such that each node $s'$
with 
$\rma(s) \prec s' \preceq s$ 
is the largest in the $\sib$ ordering. This naturally
defines a string of labels, by collecting all labels of nodes between
$\rma(s)$ and $s$. We denote this string by $\rms(s)$. 
For example, if $s$ is the rightmost node in the tree shown in
Fig.~\ref{fig-stream}, then $\rms(s)=aac$. 
Finally, for
each regular language $L$ over strings, we write $U_L^{\rms}(s)$ iff
$\rms(s)\in L$.

The following is due to Segoufin and the author.

\begin{proposition}
\label{luc-prop}
An $\MSO$ sentence $\phi$ over ordered unranked trees is streamable
iff it is expressible in \MSO\ over the vocabulary that includes
$\fc$, $(P_a)_{a\in \Sigma}$, and $U_L^{\rms}$, where $L$ ranges over
regular languages.
\end{proposition}

However, the decidability of checking whether an \MSO\ sentence
belongs to the fragment of Proposition \ref{luc-prop} remains open. 

Some recent results on processing queries over streaming data
(especially XPath queries) can be found in
\cite{baryosef,grohe-icalp05}.

\subsection{Future directions and open questions} 

\begin{enumerate}

\item This survey has concentrated primarily on Boolean and unary
  queries. While these are sufficient in many applications, there are
  formalisms that require more general $n$-ary queries. For example,
  the core expressions of XQuery can be seen as rearranging arbitrary
  $n$-tuples of nodes selected from a tree as another tree. The
  logical study of XQuery is just beginning \cite{koch-pods05}, and
  there are several papers that show how to extend results from logics
  that define Boolean and unary queries to arbitrary $n$-ary
  queries. For example, \cite{diving} does it for queries definable in 
  $\FO(\text{\rm Reg})$-like logics. Using a similar approach,
  \cite{cav-not-quite} shows how to combine temporal logics over trees to
  define $n$-ary queries. An extension of unranked tree automata to
  $n$-ary queries is presented in \cite{niehren-nary}.

\item While we have a number of logics that provide a declarative
approach to expressing properties of trees and yet match (or are close
to) the complexity of the procedural automata formalism, it is not 
really understood what causes certain logics to have such a nice
behavior. There must be some intrinsic properties of logics that lead
to good model-checking algorithms (in a way similar to, say, finite- or
tree-model properties being an explanation for decidability).

\item Closely related to the first item is the issue of 
succinctness of logics, measured as the size of formulae needed to
express certain properties. Initial investigation on the issue of
succinctness for logics on ranked trees was done in \cite{grohe-succ}
and some logics have been shown to be much more succinct than others,
but more needs to be done. In view of the standard translation between
ranked and unranked trees, it is likely that results for binary trees
will be sufficient.

\item The connection between FO, MSO, temporal logics 
and 
logics used in the programming languages and
computational linguistics 
communities must be understood. The focus was quite different, as we
mentioned earlier: for example, many questions about the complexity
and expressiveness of ambient logics are unresolved. Some very recent
results in this direction are reported in \cite{boneva-lics}. 

\item XML trees in addition to labels have data values associated with
some nodes (typically attribute values or {\tt PCDATA} values). Adding
values from a potentially infinite set and just equality over them
immediately leads to undecidable formalisms. 
This is observed, in particular, in the study of XML constraints. Some
typically considered constraints include keys and foreign keys, that
arise naturally when relational data is converted into XML. Keys say
that a certain sequence of attributes identifies a node uniquely. A
key is unary if it consists of one attribute (for example, a unique id
would be a unary key, while a pair (firstname,lastname) can be a key
consisting of two attributes). A foreign key states that a sequence of
attributes of each node labeled by $a_1$ should also occur as a sequence
of attributes of some other node labeled $a_2$.

XML specifications may consist of DTDs together with
constraints. However, their interaction could be quite complicated. In
fact, \cite{jacm-pods01} showed that it is undecidable whether a
specification that consists of a DTD
and a set of keys and foreign keys is consistent. However, if all keys
and foreign keys are unary, then consistency checking is NP-complete. 

It would be nice to find a purely logical explanation for this type of
results. Decidability restrictions studied in 
\cite{NSV-mfcs,therien} are very weak for this purpose. However, a
recent line of results shows much more promise. Consider trees
that can carry data values, and assume that we can
test them for equality, that is, we have a binary relation $\sim$ that
is true if two nodes in a tree have the same data values. Then 
$\FO^2$ over such trees with the $\sim$ relation and the successor
relation is 
decidable \cite{luc-pods06}. Here $\FO^2$ refers to \FO\ with two
variables. Notice that for expressing unary constraints two variables
suffice. It is open whether the descendant can be added while
preserving decidability; the only resolved case is that of strings,
where indeed $\FO^2$ over the successor relation, the linear ordering,
and the $\sim$ relation is decidable \cite{luc-lics06}.

\end{enumerate}

\section*{Acknowledgement}
I am grateful to Cristiana Chitic, Christoph Koch, Maarten Marx, Frank
Neven, Joachim Niehren, Gerald Penn, Thomas Schwentick, Luc Segoufin,
Anthony Widjaja To, and the referees for their comments on the
paper. I also thank Luc Segoufin for his permission to include
Proposition \ref{luc-prop} in the survey. This work was supported by
grants from NSERC and CITO.

\newcommand{\oneurl}[1]{\texttt{#1}}
%\small

\end{document}